\documentclass{aa}  

\usepackage{graphicx}
\usepackage{lscape}
\usepackage{txfonts}
\usepackage{hyperref}
\begin{document}
   \title{New and updated stellar parameters for 90 transit hosts\thanks{The data presented herein are based on observations collected at the La Silla Paranal Observatory, ESO (Chile) with the FEROS spectrograph at the 2.2-m telescope (ESO runs ID 088.C-0892, 089.C-0444, 090.C-0146) and the HARPS spectrograph at the 3.6-m telescope (ESO archive), the Paranal Observatory, ESO (Chile) with the UVES spectrograph at the VLT Kueyen telescope (ESO run ID 083.C-0174), at the Spanish Observatorio del Roque de los Muchachos of the Instituto de Astrofisica de Canarias with the FIES spectrograph at the Nordic Optical Telescope, operated on the island of La Palma jointly by Denmark, Finland, Iceland, Norway, and Sweden (program ID 40-203), and at the Observatoire de Haute-Provence (OHP, CNRS/OAMP), France with the SOPHIE spectrographs at the 1.93-m telescope (program ID 11B.DISC.SOUS).}}

   \subtitle{The effect of the surface gravity}

   \author{A. Mortier\inst{1,2},
           N.C. Santos\inst{1,2},
           S.G. Sousa\inst{1,3},
           J.M. Fernandes\inst{4},
           V.Zh. Adibekyan\inst{1},
           E. Delgado Mena\inst{1},
           M. Montalto\inst{1},
           \and
           G. Israelian\inst{3,5}
          }

   \institute{Centro de Astrof\'{\i}sica, Universidade do Porto, Rua das Estrelas, 4150-762 Porto, Portugal\\
              \email{amortier@astro.up.pt}
	      \and
	      Departamento de F\'{\i}sica e Astronomia, Faculdade de Ci\^encias, Universidade do Porto, Portugal	      
	\and
	      Instituto de Astrof\'{\i}sica de Canarias, 38200 La Laguna, Tenerife, Spain 
	\and
                      CGUC, Departement of Mathematics and Astronomical Observatory, University of Coimbra, Portugal
	\and
	      Departemento de Astrof\'{\i}sica, Universidad de La Laguna, E-38206 La Laguna, Tenerife, Spain
	      }

   \date{Received ...; Accepted ...}

% \abstract{}{}{}{}{} 
% 5 {} token are mandatory
 
  \abstract
  % context heading (optional), leave it empty if necessary  
   {Precise stellar parameters are crucial in exoplanet research for correctly determining of the planetary parameters. For stars hosting a transiting planet, determining of the planetary mass and radius depends on the stellar mass and radius, which in turn depend on the atmospheric stellar parameters. Different methods can provide different results, which leads to different planet characteristics. }%Spectroscopic surface gravities have shown to be poorly constrained, but the photometry of the transiting planet can provide an independent measurement of the surface gravity.}
   {In this paper, we use a uniform method to spectroscopically derive stellar atmospheric parameters, chemical abundances, stellar masses, and stellar radii for a sample of 90 transit hosts. Surface gravities are also derived photometrically using the stellar density as derived from the light curve. We study the effect of using these different surface gravities on the determination of the chemical abundances and the stellar mass and radius.}
   {A spectroscopic analysis based on Kurucz models in LTE was performed through the MOOG code to derive the atmospheric parameters and the chemical abundances. The photometric surface gravity was determined through isochrone fitting and the use of the stellar density, directly determined from the light curve. Stellar masses and radii are determined through calibration formulae.}
   {Spectroscopic and photometric surface gravities differ, but this has very little effect on the precise determination of the stellar mass in our spectroscopic analysis. The stellar radius, and hence the planetary radius, is most affected by the surface gravity discrepancies. For the chemical abundances, the difference is, as expected, only noticable for the abundances derived from analyzing of lines of ionized species.}
%   {Caution should be placed when determining planetary radii, since the effect of the surface gravity can have different radius determinations up to 50\%.}
    {}

   \authorrunning{Mortier, A. et al.}
   \maketitle
%
%________________________________________________________________

\section{Introduction}

Since the discovery of the first extrasolar planet around a solar-like star in 1995 \citep[51 Peg b, ][]{Mayor95}, the search for extrasolar planetary systems has accelerated. Today, more than 900 planets have been announced\footnote{\url{www.exoplanet.eu} for an updated number}. Most of them were detected using the radial velocity technique, but in the past few years, the photometric transit technique has started to produce a large number of results thanks to big space and ground missions, such as Kepler, CoRoT, and WASP \citep[e.g. ][]{And10,Leg09,Bat13}. Over 200 stars have been confirmed so far to be transited by one or more planets.

Transiting extrasolar planets have been found orbiting different types of stars, and the planets themselves also turn out to be very diverse. The large number of discoveries combined with this diversity in the planets and their hosts opens the possibility of comparing the observed properties with those predicted by theoretical models \citep[e.g. ][]{Mig11,Mor12,Mor12b}. This will put constraints on the models and help in our understanding of planet formation.

However, derivation of the planetary properties (mass, radius, and mean density) depends considerably on the deduced parameters for the stellar hosts \citep[e.g. ][]{Bou04,Tor12}. For a transiting planet, analysis of the lightcurve only determines the planetary radius relative to the stellar radius ($R_p/R_{\ast}$). Also, the planetary mass depends on the stellar mass ($M_p \propto M_{\ast}^{2/3}$), as derived from the radial velocity curve. Deriving the stellar radius and mass in turn depends on the effective temperature, surface gravity, and the metallicity of the star. 

It is thus extremely important to use high-quality data to refine the values for these stellar properties to obtain more precise stellar masses and radii and therefore more precise planetary masses and radii. Furthermore, to minimize the errors, a uniform analysis is required \citep{Tor08,Tor12} to guarantee the best possible homogeneity in the results. Using different methods to derive stellar properties leads to discrepancies in the results, which in turn leads to less significance for the statistical analyses of the data. If, for example, stellar radii were underestimated, the planetary radii would be underestimated. The occurrence rate of small planets \citep[e.g. ][]{Dre13} in our Galaxy will be affected by this underestimation. 

By homogeneously deriving precise stellar parameters for planet hosts, we gain more than just improving the planetary parameters. Observational and theoretical evidence shows that the presence of a planet seems to depend on several stellar properties, such as mass and metallicity \citep{Udry07,Sou11b,Mayor11,ME13}.
Several other correlations have come to light with the increasing discoveries of extrasolar planets, like the radius anomaly. There is evidence for a possible correlation between planetary effective temperature, metallicity, and the radius anomaly (between the observed radius and the one expected from planetary models) for giant planets \citep{Gui06,Bur07,Lau11}. According to basic core accretion theory, higher metallicities lead to larger planet cores, and such planets would have smaller radii than similar-mass planets with small or no cores. If this is true, the metallicity should be a determining factor in the observed radius anomaly and in the chemical composition and structure of the planets. Precise metallicities are thus crucial for understanding these possible correlations.

In this paper, we homogeneously derive stellar parameters and chemical abundances for a large sample of transit hosts. We also take a closer look at the surface gravity and its effect on the stellar mass and radius determinations. In Section \ref{Dat}, we present the sample that has been used and the observations. Section \ref{Spec} describes the spectroscopic analysis that was performed, as well as the results. Section \ref{logg} handles the effect of the surface gravity on the stellar mass and radius and on the chemical abundances. In Section \ref{Lit}, we compare our results with the literature. We discuss in Section \ref{Disc} and conclude in Section \ref{Con}.

\begin{table}
\caption{Observation log of the transit hosts analyzed previously with the same method used in this work.}
\label{TabLog2}
\centering
\begin{tabular}{lll}
\hline\hline
Name & Instrument & Reference \\
\hline
\object{HAT-P-1} & SARG      & 1 \\
\object{HAT-P-4} & SOPHIE         & 1 \\
\object{HAT-P-6} & SOPHIE         & 1 \\
\object{HAT-P-7} & SOPHIE         & 1 \\
\object{HD149026} & SARG      & 1 \\
\object{HD17156} & SOPHIE         & 1 \\
\object{HD189733} & CORALIE  & 2 \\
\object{HD209458} & HARPS    & 3 \\
\object{HD80606} & UES                & 4 \\
\object{HD97658} & UVES & 5 \\
\object{Kepler-17} & SOPHIE & 6 \\
\object{Kepler-21} & NARVAL & 7 \\
\object{KOI-135} & SOPHIE & 6 \\
\object{KOI-204} & SOPHIE & 6 \\
\object{OGLE-TR-10} & UVES     & 8 \\
\object{OGLE-TR-111} & UVES     & 8 \\
\object{OGLE-TR-113} & UVES     & 8 \\
\object{OGLE-TR-132} & UVES     & 9 \\
\object{OGLE-TR-182} & UVES     & 10 \\
\object{OGLE-TR-211} & UVES     & 11 \\
\object{OGLE-TR-56} & UVES     & 8 \\
\object{TrES-1} & UVES     & 8 \\
\object{TrES-2} & SARG      & 1 \\
\object{TrES-3} & SARG      & 1 \\
\object{TrES-4} & SOPHIE         & 1 \\
\object{WASP-13} & HIRES & 12 \\
\object{XO-1} & SARG      & 1 \\
\object{XO-2} & SOPHIE & 1 \\

\hline
\end{tabular}
\tablebib{(1)~\citet{Amm09};
(2) \citet{Sou06}; (3) \citet{Sou08}; (4) \citet{San04};
(5) \citet{Sou13}; (6) \citet{Bono12}; (7) \citet{Mol13};
(8) \citet{San06}; (9) \citet{Gil07}; (10) \citet{Pon08};
(11) \citet{Uda08}; (12) \citet{Gom13}.
}
\end{table}

\begin{table}
\caption{Observation log of the transit hosts analyzed in this work.}
\label{TabLog1}
\centering
\begin{tabular}{p{5cm}c}
\hline\hline
Name & Instrument \\
\hline
\object{HAT-P-17}, \object{HAT-P-20}, \object{HAT-P-26}, & \\
\object{HAT-P-30}, \object{HAT-P-35}, \object{WASP-12}, & \\
\object{WASP-18}, \object{WASP-21}, \object{WASP-26}, & \\
\object{WASP-29}, \object{WASP-32}, \object{WASP-34}, & \\
\object{WASP-35}, \object{WASP-42}, \object{WASP-45}, & FEROS\\
\object{WASP-47}, \object{WASP-50}, \object{WASP-54}, & \\
\object{WASP-55}, \object{WASP-56}, \object{WASP-62}, & \\ 
\object{WASP-63}, \object{WASP-66}, \object{WASP-67}, & \\
\object{WASP-71}, \object{WASP-77A}, \object{WASP-78}, & \\
\object{WASP-79}, \object{WASP-8} & \\
 & \\
\object{HAT-P-8} & FIES \\
 & \\
\object{CoRoT-1}, \object{CoRoT-10}, \object{CoRoT-12}, & \\
\object{CoRoT-4}, \object{CoRoT-5}, \object{CoRoT-7}, & \\
\object{CoRoT-8}, \object{CoRoT-9}, \object{HAT-P-27}, & \\
\object{WASP-15}, \object{WASP-16}, \object{WASP-17}, & HARPS \\
\object{WASP-19}, \object{WASP-22}, \object{WASP-23}, & \\
\object{WASP-24}, \object{WASP-25}, \object{WASP-28}, & \\
\object{WASP-31}, \object{WASP-36}, \object{WASP-38}, & \\
\object{WASP-41}, \object{WASP-6} & \\
 & \\
\object{HAT-P-11} & SOPHIE \\
 & \\
\object{CoRoT-2}, \object{WASP-1}, \object{WASP-10}, & \\ 
\object{WASP-11}, \object{WASP-2}, \object{WASP-4}, & UVES \\
\object{WASP-5}, \object{WASP-7} & \\

\hline
\end{tabular}
\end{table}

%__________________________________________________________________

\section{The sample}\label{Dat}

For this analysis, we used a sample of 90 stars. All these stars are of spectral type F, G or K and are known to be orbited by a transiting planet (according to the online catalog \url{www.exoplanet.eu}). From this sample, 28 stars were previously analyzed and published by members of our team. The references can be found in Table \ref{TabLog2}. For the 62 remaining stars, we gathered spectra through observations made by our team and the use of the ESO archive (see Table \ref{TabLog1}).

In total, ten different high-resolution spectrographs were used (see Table \ref{TabINS}): UVES (VLT Kueyen telescope, Paranal, Chile), FEROS (2.2m ESO/MPI telescope, La Silla, Chile), HARPS (3.6m ESO telescope, La Silla, Chile), CORALIE (1.2m Swiss telescope, La Silla, Chile), SOPHIE (1.93m telescope, OHP, France), SARG (TNG Telescope, La Palma, Spain), FIES (Nordic Optical Telescope, La Palma, Spain), NARVAL (2m T\'elescope Bernard Lyot, OPM, France), HIRES (Keck-I, Paranal, Chile) and UES (William Herschel Telescope, La Palma, Spain). The spectra were reduced using the available pipelines and IRAF\footnote{IRAF is distributed by National Optical Astronomy Observatories, operated by the Association of Universities for Research in Astronomy, Inc., under contract with the National Science Foundation, USA.}. The spectra were corrected for radial velocity with the IRAF task \texttt{DOPCOR}, to put the lines in their rest frame. To correct for this, we used the very recognizable \ion{Fe}{i} line at 6705.11\AA. Individual exposures of multiple observed stars with the same instrument were added using the task \texttt{SCOMBINE} in IRAF. The data logs can be found in Table \ref{TabLog2} and \ref{TabLog1}.

\begin{table}
\caption{Spectrograph details: resolving power and spectral ranges.}
\label{TabINS}
\centering
\begin{tabular}{cccc}
\hline\hline
Instrument & Resolving power & Spectral range & Stars \\
 & $\lambda/\Delta\lambda$ & \AA & \\
\hline
CORALIE & 50000 & 3800 - 6800 & 1 \\
FEROS & 48000 & 3600 - 9200 & 29 \\
FIES & 67000 & 3700 - 7300 & 1 \\
HARPS & 100000 & 3800 - 7000 & 24 \\
HIRES & 72000 & 4800 - 8000 & 1 \\
NARVAL & 75000 & 3700 - 10500 & 2 \\
SOPHIE & 75000 & 3820 - 6920 & 10 \\
SARG & 57000 - 86000 & 5100 - 10100 & 5 \\
UES & 55000 & 4000 - 10000 & 1 \\
UVES & 110000 & 3000 - 6800 & 16 \\
\hline
\end{tabular}
\end{table}

So far, 234 FGK planet hosts have been discovered, that are orbited by at least one transiting planet\footnote{according to \url{exoplanet.eu} on 8 July 2013}. With our sample of 90 stars, we thus analyze $\sim40\%$ of all known transit hosts. Our analysis requires high-resolution and high signal-to-noise (S/N) spectra, which is, unfortunately, not always easy to acquire for these transit hosts, since they are, on average, fainter than radial velocity hosts. Our spectra have a S/N between 100 and 300.

\section{Spectroscopic analysis}\label{Spec}

\subsection{Atmospheric parameters}

From the spectra, we derived the following atmospheric stellar parameters: the effective temperature T$_{eff}$, the surface gravity $\log g$, the metallicity [Fe/H], and the microturbulence $\xi$. The procedure we followed is described in \citet{San04} and is based on the equivalent widths of \ion{Fe}{i} and \ion{Fe}{ii} lines and on iron excitation and ionization equilibrium, assumed in local thermodynamic equilibrium (LTE). The 2010 version of MOOG\footnote{\url{http://www.as.utexas.edu/~chris/moog.html}} \citep{Sne73}, a grid of ATLAS plane-parallel model atmospheres \citep{Kur93}, and the iron linelist of \citet{Sou08} are therefore used.

To measure the equivalent widths of the iron lines, the code ARES was used \citep[Automatic Routine for line Equivalent widths in stellar Spectra -][]{Sou07}. The input parameters for ARES, are the same as in \citet{Sou08}, except for the \emph{ rejt} parameter, which determines the calibration of the continuum position. Since this parameter strongly depends on the S/N of the spectra, different values are needed for each spectrum. A uniform S/N value is derived for the spectra with the IRAF routine \texttt{BPLOT}. Therefore, three spectral regions are used: [5744\AA, 5747\AA], [6047\AA, 6053\AA], and [6068\AA, 6076\AA]. 

Then, the \emph{ rejt} parameter was set by eye for a couple of spectra with different S/N (representable for the whole sample). Afterwards, all the \emph{ rejt} parameters were derived by a simple interpolation of these values. This method ensures uniform use of the \emph{ rejt} parameter, since we otherwise do not have access to a uniform source for the S/N through the headers of the spectra as in \citet{Sou08}. The dependence of the \emph{ rejt} parameter on the S/N is the same as in \citet{ME13b}.

For cool stars, the results from using the linelist from \citet{Sou08} have shown to be unsatisfactory. The derived temperatures were higher than values from other methods, like the InfraRed flux Method \citep{Cas06}. Therefore a new linelist was built, specifically for these cooler stars \citep{Tsa13}, based on the linelist of \citet{Sou08}. Only weak and isolated lines were left, since blending effects play a huge role in cool stars. \citet{Tsa13} show that their new results are in very good agreement with the results from the InfraRed flux Method (IRFM). For the 13 stars in our sample with temperatures lower than 5200\,K, as obtained with the \citet{Sou08} linelist, we rederived the parameters with this new linelist from \citet{Tsa13}. %As a reference, we provide the results from using the \citet{Sou08} linelist in Table \ref{TabParOld}.
All atmospheric parameters can be found in Table \ref{TabPar}.

%\begin{table*}
%\caption{Stellar parameters of the cool ($T<5200$\,K) transit hosts, using the linelist from \citet{Sou08}, or the smaller linelist of \citet{San04}.}
%\label{TabParOld}
%\centering
%\begin{tabular}{cccccc}
%\hline\hline
%Name & T$_{eff}$ & $\log g_{spec}$ & [Fe/H] & $\xi$ & Reference \\
% & (K) & (dex) & (dex) & (km s$^{-1}$) & \\
%\hline
%\input{TableParOld.tex}
%\hline
%\end{tabular}
%\end{table*}

\begin{longtab}
\begin{landscape}
\begin{longtable}{cccccccccccc}
\caption{\label{TabPar} Stellar parameters for the transit hosts in this sample. The last 4 columns show the surface gravity, derived from the photometric lightcurve (LC) and the mass and radius, obtained with this surface gravity. Columns 8 and 12 show the radius, obtained through the Newton's law of gravitation.}\\
\hline\hline
Name & T$_{eff}$ & $\log g_{spec}$ & [Fe/H] & $\xi$ & M$_{\ast}$ & R$_{\ast}$ & $\rho_{\ast}$ & Ref. & $\log g_{LC}$ & M$_{\ast,LC}$ & R$_{\ast,LC}$ \\
 & (K) & (dex) & (dex) & (km s$^{-1}$) & (M$_{\odot}$) & (R$_{\odot}$) & ($\rho_{\odot}$) &  & (dex) & (M$_{\odot}$) & (R$_{\odot}$) \\
\hline
\endfirsthead
\caption{continued.}\\
\hline\hline
Name & T$_{eff}$ & $\log g_{spec}$ & [Fe/H] & $\xi$ & M$_{\ast}$ & R$_{\ast}$ & $\rho_{\ast}$ & Ref. & $\log g_{LC}$ & M$_{\ast,LC}$ & R$_{\ast,LC}$ \\
 & (K) & (dex) & (dex) & (km s$^{-1}$) & (M$_{\odot}$) & (R$_{\odot}$) & ($\rho_{\odot}$) &  & (dex) & (M$_{\odot}$) & (R$_{\odot}$) \\
\hline
\endhead
\hline
\endfoot
%%
%%% Microturbulence errors!!!
%%% OGLE 111: 1.38 -> 0.30
%%%% HAT 11: 0.70 -> 0.39
\object{CoRoT-1} & 6397 $\pm$ 54 &   4.66 $\pm$   0.09 &   0.03 $\pm$   0.04 &   1.68 $\pm$   0.09 &   1.13 $\pm$   0.08 &   0.85 $\pm$   0.10  &   0.66 $\pm$   0.02 & 1 &   4.35 $\pm$   0.01 &   1.23 $\pm$   0.08 &   1.23 $\pm$   0.04 \\
\object{CoRoT-10} & 5025 $\pm$ 155 &   4.47 $\pm$   0.31 &   0.06 $\pm$   0.09 &   1.26 $\pm$   0.34 &   0.80 $\pm$   0.10 &   0.89 $\pm$   0.41  &   2.20 $\pm$   0.47 & 1 &   4.61 $\pm$   0.02 &   0.77 $\pm$   0.07 &   0.74 $\pm$   0.04 \\
\object{CoRoT-12} & 5715 $\pm$ 208 &   4.66 $\pm$   0.22 &   0.17 $\pm$   0.14 &   1.07 $\pm$   0.31 &   0.97 $\pm$   0.10 &   0.79 $\pm$   0.23  &   0.89 $\pm$   0.08 & 1 &   4.41 $\pm$   0.02 &   1.01 $\pm$   0.10 &   1.06 $\pm$   0.06 \\
\object{CoRoT-2} & 5697 $\pm$ 97 &   4.73 $\pm$   0.17 &  -0.09 $\pm$   0.07 &   1.64 $\pm$   0.16 &   0.89 $\pm$   0.07 &   0.71 $\pm$   0.15  &   1.36 $\pm$   0.06 & 2 &   4.52 $\pm$   0.01 &   0.91 $\pm$   0.07 &   0.90 $\pm$   0.04 \\
\object{CoRoT-4} & 6344 $\pm$ 93 &   4.82 $\pm$   0.11 &   0.15 $\pm$   0.06 &   1.74 $\pm$   0.14 &   1.14 $\pm$   0.08 &   0.72 $\pm$   0.09  &   0.79 $\pm$   0.11 & 1 &   4.37 $\pm$   0.02 &   1.24 $\pm$   0.09 &   1.21 $\pm$   0.05 \\
\object{CoRoT-5} & 6240 $\pm$ 70 &   4.46 $\pm$   0.11 &   0.04 $\pm$   0.05 &   1.28 $\pm$   0.09 &   1.13 $\pm$   0.09 &   1.06 $\pm$   0.16  &   0.88 $\pm$   0.18 & 1 &   4.41 $\pm$   0.03 &   1.14 $\pm$   0.08 &   1.12 $\pm$   0.06 \\
\object{CoRoT-7} & 5288 $\pm$ 27 &   4.40 $\pm$   0.07 &   0.02 $\pm$   0.02 &   0.90 $\pm$   0.05 &   0.85 $\pm$   0.06 &   1.00 $\pm$   0.10  &   1.00 $\pm$   0.48 & 1 &   4.51 $\pm$   0.02 &   0.83 $\pm$   0.06 &   0.87 $\pm$   0.04 \\
\object{CoRoT-8} & 5143 $\pm$ 178 &   4.42 $\pm$   0.33 &   0.22 $\pm$   0.11 &   0.61 $\pm$   0.40 &   0.88 $\pm$   0.12 &   0.99 $\pm$   0.51  &   1.21 $\pm$   0.32 & 1 &   4.49 $\pm$   0.03 &   0.84 $\pm$   0.08 &   0.88 $\pm$   0.05 \\
\object{CoRoT-9} & 5613 $\pm$ 36 &   4.35 $\pm$   0.09 &  -0.02 $\pm$   0.03 &   0.90 $\pm$   0.05 &   0.94 $\pm$   0.07 &   1.11 $\pm$   0.14  &   1.16 $\pm$   0.24 & 1 &   4.47 $\pm$   0.04 &   0.91 $\pm$   0.07 &   0.96 $\pm$   0.05 \\
\object{HAT-P-1} & 6076 $\pm$ 27 &   4.47 $\pm$   0.07 &   0.21 $\pm$   0.03 &   1.17 $\pm$   0.05 &   1.12 $\pm$   0.08 &   1.04 $\pm$   0.10  &   0.82 $\pm$   0.07 & 3 &   4.40 $\pm$   0.01 &   1.15 $\pm$   0.08 &   1.13 $\pm$   0.04 \\
\object{HAT-P-11} & 4624 $\pm$ 225 &   4.15 $\pm$   0.59 &   0.26 $\pm$   0.08 &   0.39 $\pm$   0.39 &   0.85 $\pm$   0.27 &   1.34 $\pm$   1.34  &   2.42 $\pm$   0.10 & 1 &   4.65 $\pm$   0.01 &   0.72 $\pm$   0.08 &   0.67 $\pm$   0.04 \\
\object{HAT-P-17} & 5332 $\pm$ 55 &   4.45 $\pm$   0.13 &   0.05 $\pm$   0.03 &   0.82 $\pm$   0.10 &   0.86 $\pm$   0.07 &   0.95 $\pm$   0.17  &   1.46 $\pm$   0.09 & 4 &   4.52 $\pm$   0.02 &   0.85 $\pm$   0.06 &   0.87 $\pm$   0.03 \\
\object{HAT-P-20} & 4502 $\pm$ 188 &   4.32 $\pm$   0.60 &   0.12 $\pm$   0.15 &   0.73 $\pm$   0.60 &   0.76 $\pm$   0.20 &   1.03 $\pm$   1.03  &   2.27 $\pm$   0.18 & 5 &   4.63 $\pm$   0.01 &   0.69 $\pm$   0.07 &   0.66 $\pm$   0.04 \\
\object{HAT-P-26} & 5011 $\pm$ 55 &   4.31 $\pm$   0.17 &   0.01 $\pm$   0.04 &   0.48 $\pm$   0.16 &   0.81 $\pm$   0.08 &   1.07 $\pm$   0.27  &   1.69 $\pm$   0.32 & 6 &   4.56 $\pm$   0.02 &   0.77 $\pm$   0.06 &   0.78 $\pm$   0.03 \\
\object{HAT-P-27} & 5316 $\pm$ 55 &   4.48 $\pm$   0.10 &   0.30 $\pm$   0.03 &   0.82 $\pm$   0.09 &   0.90 $\pm$   0.07 &   0.93 $\pm$   0.12  &   1.32 $\pm$   0.19 & 7 &   4.51 $\pm$   0.03 &   0.89 $\pm$   0.07 &   0.89 $\pm$   0.04 \\
\object{HAT-P-30} & 6338 $\pm$ 42 &   4.52 $\pm$   0.06 &   0.12 $\pm$   0.03 &   1.40 $\pm$   0.05 &   1.17 $\pm$   0.08 &   1.00 $\pm$   0.08  &   0.70 $\pm$   0.07 & 8 &   4.36 $\pm$   0.01 &   1.23 $\pm$   0.08 &   1.22 $\pm$   0.05 \\
\object{HAT-P-35} & 6178 $\pm$ 45 &   4.40 $\pm$   0.09 &   0.12 $\pm$   0.03 &   1.34 $\pm$   0.06 &   1.16 $\pm$   0.08 &   1.14 $\pm$   0.14  &   0.42 $\pm$   0.06 & 9 &   4.22 $\pm$   0.03 &   1.25 $\pm$   0.08 &   1.44 $\pm$   0.08 \\
\object{HAT-P-4} & 6054 $\pm$ 60 &   4.17 $\pm$   0.28 &   0.35 $\pm$   0.08 &   1.59 $\pm$   0.09 &   1.37 $\pm$   0.18 &   1.57 $\pm$   0.71  &   0.31 $\pm$   0.03 & 1 &   4.14 $\pm$   0.02 &   1.35 $\pm$   0.09 &   1.60 $\pm$   0.07 \\
\object{HAT-P-6} & 6855 $\pm$ 111 &   4.69 $\pm$   0.20 &  -0.08 $\pm$   0.11 &   2.85 $\pm$   1.15 &   1.26 $\pm$   0.10 &   0.86 $\pm$   0.22  &   0.37 $\pm$   0.04 & 2 &   4.20 $\pm$   0.02 &   1.47 $\pm$   0.11 &   1.57 $\pm$   0.08 \\
\object{HAT-P-7} & 6525 $\pm$ 61 &   4.09 $\pm$   0.08 &   0.31 $\pm$   0.07 &   1.78 $\pm$   0.14 &   1.64 $\pm$   0.11 &   1.81 $\pm$   0.22  &   0.20 $\pm$   0.01 & 1 &   4.04 $\pm$   0.01 &   1.69 $\pm$   0.10 &   1.94 $\pm$   0.07 \\
\object{HAT-P-8} & 6550 $\pm$ 61 &   4.80 $\pm$   0.08 &   0.07 $\pm$   0.04 &   1.93 $\pm$   0.09 &   1.18 $\pm$   0.08 &   0.74 $\pm$   0.07  &   0.37 $\pm$   0.04 & 10 &   4.19 $\pm$   0.03 &   1.41 $\pm$   0.09 &   1.56 $\pm$   0.08 \\
\object{HD149026} & 6162 $\pm$ 41 &   4.37 $\pm$   0.10 &   0.36 $\pm$   0.05 &   1.41 $\pm$   0.07 &   1.26 $\pm$   0.09 &   1.21 $\pm$   0.17  &   0.59 $\pm$   0.11 & 3 &   4.33 $\pm$   0.03 &   1.28 $\pm$   0.09 &   1.26 $\pm$   0.07 \\
\object{HD17156} & 6084 $\pm$ 29 &   4.33 $\pm$   0.05 &   0.23 $\pm$   0.04 &   1.47 $\pm$   0.05 &   1.19 $\pm$   0.08 &   1.24 $\pm$   0.09  &   0.40 $\pm$   0.02 & 1 &   4.21 $\pm$   0.01 &   1.26 $\pm$   0.08 &   1.46 $\pm$   0.05 \\
\object{HD189733} & 5109 $\pm$ 146 &   4.69 $\pm$   0.28 &   0.03 $\pm$   0.08 &   0.78 $\pm$   0.33 &   0.79 $\pm$   0.08 &   0.70 $\pm$   0.26  &   1.98 $\pm$   0.17 & 3 &   4.60 $\pm$   0.01 &   0.78 $\pm$   0.07 &   0.76 $\pm$   0.03 \\
\object{HD209458} & 6118 $\pm$ 25 &   4.50 $\pm$   0.04 &   0.03 $\pm$   0.02 &   1.21 $\pm$   0.03 &   1.07 $\pm$   0.07 &   0.99 $\pm$   0.06  &   0.73 $\pm$   0.01 & 3 &   4.36 $\pm$   0.01 &   1.12 $\pm$   0.08 &   1.18 $\pm$   0.04 \\
\object{HD80606} & 5574 $\pm$ 72 &   4.46 $\pm$   0.20 &   0.32 $\pm$   0.09 &   1.14 $\pm$   0.09 &   1.00 $\pm$   0.09 &   1.00 $\pm$   0.28  &   0.91 $\pm$   0.06 & 1 &   4.42 $\pm$   0.02 &   1.00 $\pm$   0.08 &   1.05 $\pm$   0.04 \\
\object{HD97658} & 5137 $\pm$ 36 &   4.47 $\pm$   0.09 &  -0.35 $\pm$   0.02 &   0.63 $\pm$   0.08 &   0.75 $\pm$   0.06 &   0.86 $\pm$   0.11  &   1.38 $\pm$   0.45 & 11 &   4.59 $\pm$   0.01 &   0.74 $\pm$   0.05 &   0.74 $\pm$   0.03 \\
\object{Kepler-17} & 5781 $\pm$ 85 &   4.53 $\pm$   0.12 &   0.26 $\pm$   0.10 &   1.73 $\pm$   0.14 &   1.03 $\pm$   0.08 &   0.94 $\pm$   0.15  &   1.12 $\pm$   0.02 & 2 &   4.48 $\pm$   0.01 &   1.03 $\pm$   0.08 &   0.99 $\pm$   0.04 \\
\object{Kepler-21} & 6409 $\pm$ 44 &   4.43 $\pm$   0.06 &  -0.03 $\pm$   0.03 &   1.86 $\pm$   0.07 &   1.17 $\pm$   0.08 &   1.11 $\pm$   0.09  &   0.20 $\pm$   0.01 & 12 &   4.03 $\pm$   0.05 &   1.44 $\pm$   0.10 &   1.88 $\pm$   0.15 \\
\object{KOI-135} & 6041 $\pm$ 143 &   4.26 $\pm$   0.05 &   0.33 $\pm$   0.11 &   1.85 $\pm$   0.26 &   1.25 $\pm$   0.10 &   1.36 $\pm$   0.11  &   0.52 $\pm$   0.02 & 2 &   4.28 $\pm$   0.01 &   1.24 $\pm$   0.10 &   1.33 $\pm$   0.06 \\
\object{KOI-204} & 5757 $\pm$ 134 &   4.15 $\pm$   0.06 &   0.26 $\pm$   0.10 &   1.75 $\pm$   0.19 &   1.17 $\pm$   0.10 &   1.51 $\pm$   0.14  &   0.39 $\pm$   0.07 & 2 &   4.17 $\pm$   0.05 &   1.16 $\pm$   0.10 &   1.48 $\pm$   0.11 \\
\object{OGLE-TR-10} & 6075 $\pm$ 86 &   4.54 $\pm$   0.15 &   0.28 $\pm$   0.10 &   1.45 $\pm$   0.14 &   1.14 $\pm$   0.09 &   0.97 $\pm$   0.19  &   0.36 $\pm$   0.06 & 3 &   4.18 $\pm$   0.04 &   1.30 $\pm$   0.10 &   1.52 $\pm$   0.10 \\
\object{OGLE-TR-111} & 4800 $\pm$ 177 &   4.24 $\pm$   0.46 &   0.22 $\pm$   0.15 &   0.30 $\pm$   0.30 &   0.85 $\pm$   0.19 &   1.19 $\pm$   0.97  &   1.52 $\pm$   0.10 & 2 &   4.54 $\pm$   0.01 &   0.76 $\pm$   0.08 &   0.79 $\pm$   0.04 \\
\object{OGLE-TR-113} & 4781 $\pm$ 166 &   4.31 $\pm$   0.41 &   0.03 $\pm$   0.06 &   1.24 $\pm$   0.29 &   0.79 $\pm$   0.14 &   1.06 $\pm$   0.73  &   1.68 $\pm$   0.06 & 2 &   4.56 $\pm$   0.01 &   0.73 $\pm$   0.07 &   0.75 $\pm$   0.04 \\
\object{OGLE-TR-132} & 6210 $\pm$ 59 &   4.51 $\pm$   0.27 &   0.37 $\pm$   0.07 &   1.23 $\pm$   0.09 &   1.26 $\pm$   0.12 &   1.04 $\pm$   0.40  &   0.54 $\pm$   0.06 & 2 &   4.30 $\pm$   0.03 &   1.32 $\pm$   0.09 &   1.33 $\pm$   0.06 \\
\object{OGLE-TR-182} & 5924 $\pm$ 64 &   4.47 $\pm$   0.18 &   0.37 $\pm$   0.08 &   0.91 $\pm$   0.09 &   1.14 $\pm$   0.10 &   1.04 $\pm$   0.26  &   0.33 $\pm$   0.10 & 3 &   4.15 $\pm$   0.07 &   1.29 $\pm$   0.10 &   1.57 $\pm$   0.16 \\
\object{OGLE-TR-211} & 6325 $\pm$ 91 &   4.22 $\pm$   0.17 &   0.11 $\pm$   0.10 &   1.63 $\pm$   0.21 &   1.32 $\pm$   0.13 &   1.47 $\pm$   0.37  &   0.34 $\pm$   0.08 & 3 &   4.17 $\pm$   0.05 &   1.34 $\pm$   0.10 &   1.56 $\pm$   0.13 \\
\object{OGLE-TR-56} & 6119 $\pm$ 62 &   4.21 $\pm$   0.19 &   0.25 $\pm$   0.08 &   1.48 $\pm$   0.11 &   1.30 $\pm$   0.13 &   1.48 $\pm$   0.43  &   0.26 $\pm$   0.01 & 2 &   4.09 $\pm$   0.01 &   1.37 $\pm$   0.09 &   1.71 $\pm$   0.07 \\
\object{TrES-1} & 5226 $\pm$ 38 &   4.40 $\pm$   0.10 &   0.06 $\pm$   0.05 &   0.90 $\pm$   0.05 &   0.85 $\pm$   0.07 &   0.99 $\pm$   0.14  &   1.63 $\pm$   0.09 & 3 &   4.57 $\pm$   0.01 &   0.82 $\pm$   0.06 &   0.80 $\pm$   0.03 \\
\object{TrES-2} & 5795 $\pm$ 73 &   4.30 $\pm$   0.13 &   0.06 $\pm$   0.08 &   0.79 $\pm$   0.12 &   1.04 $\pm$   0.09 &   1.23 $\pm$   0.23  &   1.10 $\pm$   0.01 & 1 &   4.47 $\pm$   0.01 &   0.98 $\pm$   0.07 &   0.99 $\pm$   0.04 \\
\object{TrES-3} & 5502 $\pm$ 157 &   4.44 $\pm$   0.22 &  -0.10 $\pm$   0.19 &   1.00 $\pm$   0.30 &   0.88 $\pm$   0.10 &   0.97 $\pm$   0.31  &   1.65 $\pm$   0.04 & 1 &   4.57 $\pm$   0.03 &   0.85 $\pm$   0.09 &   0.83 $\pm$   0.05 \\
\object{TrES-4} & 6293 $\pm$ 96 &   4.20 $\pm$   0.27 &   0.34 $\pm$   0.10 &   2.01 $\pm$   0.17 &   1.46 $\pm$   0.18 &   1.55 $\pm$   0.67  &   0.22 $\pm$   0.03 & 2 &   4.06 $\pm$   0.03 &   1.55 $\pm$   0.11 &   1.85 $\pm$   0.10 \\
\object{WASP-1} & 6252 $\pm$ 45 &   4.32 $\pm$   0.05 &   0.23 $\pm$   0.03 &   1.42 $\pm$   0.05 &   1.27 $\pm$   0.09 &   1.28 $\pm$   0.10  &   0.40 $\pm$   0.05 & 2 &   4.23 $\pm$   0.03 &   1.32 $\pm$   0.09 &   1.45 $\pm$   0.07 \\
\object{WASP-10} & 4645 $\pm$ 125 &   4.27 $\pm$   0.39 &   0.04 $\pm$   0.05 &   0.58 $\pm$   0.47 &   0.76 $\pm$   0.13 &   1.08 $\pm$   0.71  &   2.16 $\pm$   0.31 & 3 &   4.61 $\pm$   0.02 &   0.70 $\pm$   0.06 &   0.69 $\pm$   0.03 \\
\object{WASP-11} & 4881 $\pm$ 125 &   4.44 $\pm$   0.31 &   0.01 $\pm$   0.05 &   0.64 $\pm$   0.24 &   0.77 $\pm$   0.09 &   0.90 $\pm$   0.42  &   2.12 $\pm$   0.46 & 13 &   4.63 $\pm$   0.02 &   0.74 $\pm$   0.06 &   0.71 $\pm$   0.03 \\
\object{WASP-12} & 6313 $\pm$ 52 &   4.37 $\pm$   0.12 &   0.21 $\pm$   0.04 &   1.65 $\pm$   0.07 &   1.26 $\pm$   0.10 &   1.21 $\pm$   0.21  &   0.22 $\pm$   0.02 & 2 &   4.05 $\pm$   0.02 &   1.49 $\pm$   0.10 &   1.85 $\pm$   0.08 \\
\object{WASP-13} & 6025 $\pm$ 21 &   4.19 $\pm$   0.03 &   0.11 $\pm$   0.05 &   1.28 $\pm$   0.10 &   1.20 $\pm$   0.08 &   1.46 $\pm$   0.08  &   0.13 $\pm$   0.02 & 2 &   3.90 $\pm$   0.03 &   1.45 $\pm$   0.10 &   2.20 $\pm$   0.13 \\
\object{WASP-15} & 6573 $\pm$ 70 &   4.79 $\pm$   0.08 &   0.09 $\pm$   0.04 &   1.72 $\pm$   0.09 &   1.20 $\pm$   0.08 &   0.76 $\pm$   0.07  &   0.39 $\pm$   0.03 & 14 &   4.22 $\pm$   0.02 &   1.41 $\pm$   0.09 &   1.50 $\pm$   0.06 \\
\object{WASP-16} & 5726 $\pm$ 22 &   4.34 $\pm$   0.05 &   0.13 $\pm$   0.02 &   0.97 $\pm$   0.03 &   1.02 $\pm$   0.07 &   1.16 $\pm$   0.09  &   1.21 $\pm$   0.15 & 15 &   4.49 $\pm$   0.02 &   0.98 $\pm$   0.07 &   0.95 $\pm$   0.04 \\
\object{WASP-17} & 6794 $\pm$ 83 &   4.83 $\pm$   0.09 &  -0.12 $\pm$   0.05 &   2.57 $\pm$   0.22 &   1.20 $\pm$   0.08 &   0.73 $\pm$   0.08  &   0.32 $\pm$   0.01 & 2 &   4.16 $\pm$   0.01 &   1.45 $\pm$   0.10 &   1.64 $\pm$   0.06 \\
\object{WASP-18} & 6526 $\pm$ 69 &   4.73 $\pm$   0.08 &   0.19 $\pm$   0.05 &   1.83 $\pm$   0.10 &   1.23 $\pm$   0.08 &   0.81 $\pm$   0.08  &   0.69 $\pm$   0.06 & 2 &   4.32 $\pm$   0.03 &   1.37 $\pm$   0.09 &   1.33 $\pm$   0.07 \\
\object{WASP-19} & 5591 $\pm$ 62 &   4.46 $\pm$   0.09 &   0.26 $\pm$   0.05 &   1.23 $\pm$   0.09 &   0.98 $\pm$   0.08 &   0.99 $\pm$   0.12  &   0.99 $\pm$   0.04 & 16 &   4.44 $\pm$   0.01 &   0.98 $\pm$   0.07 &   1.02 $\pm$   0.04 \\
\object{WASP-2} & 5109 $\pm$ 72 &   4.33 $\pm$   0.14 &   0.02 $\pm$   0.05 &   0.57 $\pm$   0.12 &   0.83 $\pm$   0.07 &   1.06 $\pm$   0.21  &   1.52 $\pm$   0.07 & 2 &   4.54 $\pm$   0.01 &   0.79 $\pm$   0.06 &   0.81 $\pm$   0.03 \\
\object{WASP-21} & 5924 $\pm$ 55 &   4.39 $\pm$   0.09 &  -0.22 $\pm$   0.04 &   1.06 $\pm$   0.08 &   0.97 $\pm$   0.07 &   1.08 $\pm$   0.14  &   0.59 $\pm$   0.06 & 2 &   4.28 $\pm$   0.03 &   1.00 $\pm$   0.07 &   1.24 $\pm$   0.06 \\
\object{WASP-22} & 6153 $\pm$ 46 &   4.57 $\pm$   0.09 &   0.26 $\pm$   0.03 &   1.36 $\pm$   0.06 &   1.14 $\pm$   0.08 &   0.93 $\pm$   0.11  &   0.61 $\pm$   0.06 & 17 &   4.32 $\pm$   0.02 &   1.23 $\pm$   0.08 &   1.27 $\pm$   0.05 \\
\object{WASP-23} & 5046 $\pm$ 99 &   4.33 $\pm$   0.18 &   0.05 $\pm$   0.06 &   0.64 $\pm$   0.23 &   0.82 $\pm$   0.08 &   1.06 $\pm$   0.28  &   1.84 $\pm$   0.03 & 18 &   4.59 $\pm$   0.01 &   0.78 $\pm$   0.06 &   0.76 $\pm$   0.03 \\
\object{WASP-24} & 6297 $\pm$ 58 &   4.76 $\pm$   0.17 &   0.09 $\pm$   0.04 &   1.41 $\pm$   0.08 &   1.12 $\pm$   0.08 &   0.76 $\pm$   0.16  &   0.47 $\pm$   0.03 & 19 &   4.25 $\pm$   0.01 &   1.26 $\pm$   0.09 &   1.39 $\pm$   0.05 \\
\object{WASP-25} & 5736 $\pm$ 35 &   4.52 $\pm$   0.09 &   0.06 $\pm$   0.03 &   1.11 $\pm$   0.05 &   0.96 $\pm$   0.07 &   0.92 $\pm$   0.11  &   1.29 $\pm$   0.10 & 20 &   4.51 $\pm$   0.01 &   0.96 $\pm$   0.07 &   0.94 $\pm$   0.03 \\
\object{WASP-26} & 6034 $\pm$ 31 &   4.44 $\pm$   0.06 &   0.16 $\pm$   0.02 &   1.28 $\pm$   0.04 &   1.10 $\pm$   0.08 &   1.07 $\pm$   0.09  &   0.47 $\pm$   0.06 & 21 &   4.25 $\pm$   0.03 &   1.18 $\pm$   0.08 &   1.36 $\pm$   0.07 \\
\object{WASP-28} & 6134 $\pm$ 38 &   4.55 $\pm$   0.05 &  -0.12 $\pm$   0.03 &   1.17 $\pm$   0.06 &   1.02 $\pm$   0.07 &   0.92 $\pm$   0.06  &   0.93 $\pm$   0.13 & 22 &   4.44 $\pm$   0.03 &   1.04 $\pm$   0.07 &   1.05 $\pm$   0.06 \\
\object{WASP-29} & 5203 $\pm$ 102 &   4.93 $\pm$   0.21 &   0.17 $\pm$   0.05 &   1.77 $\pm$   0.22 &   0.83 $\pm$   0.07 &   0.55 $\pm$   0.13  &   1.56 $\pm$   0.21 & 23 &   4.55 $\pm$   0.02 &   0.83 $\pm$   0.07 &   0.83 $\pm$   0.04 \\
\object{WASP-31} & 6443 $\pm$ 75 &   4.76 $\pm$   0.09 &  -0.08 $\pm$   0.05 &   1.62 $\pm$   0.11 &   1.10 $\pm$   0.08 &   0.76 $\pm$   0.08  &   0.59 $\pm$   0.04 & 24 &   4.31 $\pm$   0.02 &   1.22 $\pm$   0.08 &   1.30 $\pm$   0.05 \\
\object{WASP-32} & 6427 $\pm$ 141 &   4.93 $\pm$   0.08 &   0.28 $\pm$   0.10 &   1.20 $\pm$   0.21 &   1.23 $\pm$   0.10 &   0.65 $\pm$   0.06  &   0.80 $\pm$   0.10 & 25 &   4.32 $\pm$   0.03 &   1.37 $\pm$   0.10 &   1.32 $\pm$   0.07 \\
\object{WASP-34} & 5704 $\pm$ 26 &   4.35 $\pm$   0.05 &   0.08 $\pm$   0.02 &   0.97 $\pm$   0.03 &   1.00 $\pm$   0.07 &   1.13 $\pm$   0.08  &   0.83 $\pm$   0.21 & 26 &   4.37 $\pm$   0.05 &   0.99 $\pm$   0.07 &   1.11 $\pm$   0.08 \\
\object{WASP-35} & 6072 $\pm$ 62 &   4.69 $\pm$   0.13 &  -0.05 $\pm$   0.05 &   1.26 $\pm$   0.09 &   1.00 $\pm$   0.07 &   0.79 $\pm$   0.13  &   0.83 $\pm$   0.07 & 27 &   4.39 $\pm$   0.02 &   1.06 $\pm$   0.08 &   1.12 $\pm$   0.05 \\
\object{WASP-36} & 5928 $\pm$ 59 &   4.51 $\pm$   0.09 &  -0.01 $\pm$   0.05 &   0.89 $\pm$   0.09 &   1.00 $\pm$   0.07 &   0.95 $\pm$   0.12  &   1.21 $\pm$   0.05 & 28 &   4.49 $\pm$   0.01 &   1.00 $\pm$   0.07 &   0.97 $\pm$   0.03 \\
\object{WASP-38} & 6436 $\pm$ 60 &   4.80 $\pm$   0.07 &   0.06 $\pm$   0.04 &   1.75 $\pm$   0.09 &   1.14 $\pm$   0.08 &   0.73 $\pm$   0.06  &   0.51 $\pm$   0.02 & 29 &   4.27 $\pm$   0.01 &   1.30 $\pm$   0.09 &   1.38 $\pm$   0.05 \\
\object{WASP-4} & 5513 $\pm$ 43 &   4.50 $\pm$   0.07 &   0.03 $\pm$   0.03 &   0.86 $\pm$   0.07 &   0.89 $\pm$   0.07 &   0.91 $\pm$   0.09  &   1.23 $\pm$   0.02 & 2 &   4.49 $\pm$   0.01 &   0.89 $\pm$   0.06 &   0.92 $\pm$   0.03 \\
\object{WASP-41} & 5546 $\pm$ 33 &   4.53 $\pm$   0.07 &   0.06 $\pm$   0.02 &   1.08 $\pm$   0.05 &   0.90 $\pm$   0.07 &   0.89 $\pm$   0.08  &   1.27 $\pm$   0.14 & 30 &   4.49 $\pm$   0.03 &   0.91 $\pm$   0.06 &   0.93 $\pm$   0.04 \\
\object{WASP-42} & 5315 $\pm$ 79 &   4.50 $\pm$   0.18 &   0.29 $\pm$   0.05 &   1.16 $\pm$   0.13 &   0.90 $\pm$   0.08 &   0.91 $\pm$   0.23  &   1.37 $\pm$   0.14 & 31 &   4.52 $\pm$   0.02 &   0.89 $\pm$   0.07 &   0.88 $\pm$   0.04 \\
\object{WASP-45} & 5298 $\pm$ 95 &   4.43 $\pm$   0.18 &   0.43 $\pm$   0.06 &   1.10 $\pm$   0.13 &   0.95 $\pm$   0.09 &   1.00 $\pm$   0.25  &   1.08 $\pm$   0.25 & 32 &  &  &  \\
\object{WASP-47} & 5576 $\pm$ 68 &   4.28 $\pm$   0.16 &   0.36 $\pm$   0.05 &   1.25 $\pm$   0.09 &   1.07 $\pm$   0.10 &   1.26 $\pm$   0.30  &   0.71 $\pm$   0.03 & 33 &   4.34 $\pm$   0.01 &   1.04 $\pm$   0.08 &   1.15 $\pm$   0.04 \\
\object{WASP-5} & 5785 $\pm$ 83 &   4.54 $\pm$   0.14 &   0.17 $\pm$   0.06 &   0.96 $\pm$   0.12 &   1.00 $\pm$   0.08 &   0.92 $\pm$   0.17  &   0.80 $\pm$   0.08 & 2 &   4.39 $\pm$   0.03 &   1.04 $\pm$   0.08 &   1.10 $\pm$   0.05 \\
\object{WASP-50} & 5518 $\pm$ 42 &   4.43 $\pm$   0.12 &   0.13 $\pm$   0.03 &   1.25 $\pm$   0.06 &   0.93 $\pm$   0.07 &   1.01 $\pm$   0.17  &   1.48 $\pm$   0.10 & 34 &   4.48 $\pm$   0.02 &   0.92 $\pm$   0.07 &   0.94 $\pm$   0.04 \\
\object{WASP-54} & 6296 $\pm$ 40 &   4.37 $\pm$   0.06 &   0.00 $\pm$   0.03 &   1.45 $\pm$   0.05 &   1.17 $\pm$   0.08 &   1.18 $\pm$   0.10  &   0.20 $\pm$   0.03 & 35 &   4.00 $\pm$   0.02 &   1.42 $\pm$   0.09 &   1.94 $\pm$   0.08 \\
\object{WASP-55} & 6070 $\pm$ 53 &   4.55 $\pm$   0.07 &   0.09 $\pm$   0.04 &   1.10 $\pm$   0.06 &   1.06 $\pm$   0.08 &   0.93 $\pm$   0.09  &   0.85 $\pm$   0.03 & 33 &   4.41 $\pm$   0.01 &   1.10 $\pm$   0.08 &   1.10 $\pm$   0.04 \\
\object{WASP-56} & 5797 $\pm$ 52 &   4.44 $\pm$   0.09 &   0.43 $\pm$   0.04 &   1.19 $\pm$   0.06 &   1.11 $\pm$   0.08 &   1.06 $\pm$   0.13  &   0.74 $\pm$   0.04 & 35 & & &  \\
\object{WASP-6} & 5383 $\pm$ 41 &   4.52 $\pm$   0.06 &  -0.14 $\pm$   0.03 &   0.80 $\pm$   0.07 &   0.82 $\pm$   0.06 &   0.86 $\pm$   0.07  &   1.34 $\pm$   0.11 & 36 &   4.52 $\pm$   0.01 &   0.82 $\pm$   0.06 &   0.86 $\pm$   0.03 \\
\object{WASP-62} & 6391 $\pm$ 70 &   4.73 $\pm$   0.11 &   0.24 $\pm$   0.05 &   1.50 $\pm$   0.09 &   1.20 $\pm$   0.08 &   0.80 $\pm$   0.11  &   0.59 $\pm$   0.06 & 33 &   4.33 $\pm$   0.02 &   1.33 $\pm$   0.09 &   1.29 $\pm$   0.05 \\
\object{WASP-63} & 5715 $\pm$ 60 &   4.29 $\pm$   0.10 &   0.28 $\pm$   0.05 &   1.28 $\pm$   0.07 &   1.09 $\pm$   0.09 &   1.26 $\pm$   0.18  &   0.20 $\pm$   0.02 & 33 &   4.00 $\pm$   0.02 &   1.26 $\pm$   0.09 &   1.84 $\pm$   0.09 \\
\object{WASP-66} & 7051 $\pm$ 79 &   5.00 $\pm$   0.08 &   0.05 $\pm$   0.05 &   3.07 $\pm$   0.27 &   1.39 $\pm$   0.09 &   0.64 $\pm$   0.06  &   0.24 $\pm$   0.03 & 33 &   4.10 $\pm$   0.03 &   1.75 $\pm$   0.11 &   1.84 $\pm$   0.10 \\
\object{WASP-67} & 5417 $\pm$ 85 &   4.40 $\pm$   0.16 &   0.18 $\pm$   0.06 &   1.16 $\pm$   0.12 &   0.93 $\pm$   0.08 &   1.04 $\pm$   0.23  &   1.32 $\pm$   0.15 & 33 &   4.51 $\pm$   0.02 &   0.90 $\pm$   0.07 &   0.89 $\pm$   0.04 \\
\object{WASP-7} & 6621 $\pm$ 155 &   4.62 $\pm$   0.14 &   0.12 $\pm$   0.09 &   3.00 $\pm$   0.83 &   1.26 $\pm$   0.10 &   0.92 $\pm$   0.17  &   0.41 $\pm$   0.07 & 2 &   4.22 $\pm$   0.04 &   1.45 $\pm$   0.11 &   1.52 $\pm$   0.09 \\
\object{WASP-71} & 6180 $\pm$ 52 &   4.15 $\pm$   0.06 &   0.37 $\pm$   0.04 &   1.69 $\pm$   0.06 &   1.42 $\pm$   0.10 &   1.61 $\pm$   0.15  &   0.13 $\pm$   0.02 & 37 &   3.92 $\pm$   0.03 &   1.66 $\pm$   0.10 &   2.22 $\pm$   0.12 \\
\object{WASP-77A} & 5605 $\pm$ 41 &   4.37 $\pm$   0.09 &   0.07 $\pm$   0.03 &   1.09 $\pm$   0.06 &   0.96 $\pm$   0.07 &   1.09 $\pm$   0.14  &   1.16 $\pm$   0.02 & 38 &   4.48 $\pm$   0.01 &   0.93 $\pm$   0.07 &   0.95 $\pm$   0.03 \\
\object{WASP-78} & 6291 $\pm$ 71 &   4.19 $\pm$   0.08 &  -0.07 $\pm$   0.05 &   1.63 $\pm$   0.10 &   1.24 $\pm$   0.09 &   1.49 $\pm$   0.18  &   0.12 $\pm$   0.02 & 39 &   3.89 $\pm$   0.03 &   1.51 $\pm$   0.10 &   2.25 $\pm$   0.13 \\
\object{WASP-79} & 7002 $\pm$ 162 &   4.77 $\pm$   0.14 &   0.19 $\pm$   0.10 &   2.64 $\pm$   0.24 &   1.43 $\pm$   0.11 &   0.82 $\pm$   0.14  &   0.22 $\pm$   0.03 & 39 &   4.07 $\pm$   0.03 &   1.86 $\pm$   0.13 &   1.93 $\pm$   0.11 \\
\object{WASP-8} & 5690 $\pm$ 36 &   4.42 $\pm$   0.15 &   0.29 $\pm$   0.03 &   1.25 $\pm$   0.05 &   1.04 $\pm$   0.08 &   1.07 $\pm$   0.23  &   1.22 $\pm$   0.16 & 40 &   4.48 $\pm$   0.01 &   1.01 $\pm$   0.07 &   0.98 $\pm$   0.04 \\
\object{XO-1} & 5754 $\pm$ 42 &   4.61 $\pm$   0.05 &  -0.01 $\pm$   0.05 &   1.07 $\pm$   0.09 &   0.93 $\pm$   0.07 &   0.82 $\pm$   0.06  &   1.24 $\pm$   0.08 & 3 &   4.50 $\pm$   0.01 &   0.95 $\pm$   0.07 &   0.94 $\pm$   0.04 \\
\object{XO-2} & 5350 $\pm$ 72 &   4.14 $\pm$   0.22 &   0.42 $\pm$   0.07 &   1.10 $\pm$   0.08 &   1.08 $\pm$   0.13 &   1.48 $\pm$   0.51  &   1.03 $\pm$   0.09 & 2 & & &  \\

\end{longtable}
\tablebib{(1)~\citet{South11}; 
(2) \citet{South12}; 
(3) \citet{South10}; 
(4) \citet{How12}; 
(5) \citet{Bak11}; 
(6) \citet{Hart11}; 
(7) \citet{Bek11}; 
(8) \citet{John11c}; 
(9) \citet{Bak12}; 
(10) \citet{Tod12}; 
(11) \citet{Hen11}; 
(12) \citet{Howe12}; 
(13) \citet{West09a}; 
(14) \citet{West09b}; 
(15) \citet{List09}; 
(16) \citet{Hebb10}; 
(17) \citet{Max10a}; 
(18) \citet{Tri11}; 
(19) \citet{Str10}; 
(20) \citet{Eno11a}; 
(21) \citet{Sma10}; 
(22) \citet{West13}; 
(23) \citet{Hel10}; 
(24) \citet{And11}; 
(25) \citet{Max10b}; 
(26) \citet{Sma11}; 
(27) \citet{Eno11b}; 
(28) \citet{Smi12}; 
(29) \citet{Barr11}; 
(30) \citet{Max11}; 
(31) \citet{Len12}; 
(32) \citet{And12}; 
(33) \citet{Hel12}; 
(34) \citet{Gil11}; 
(35) \citet{Fae13}; 
(36) \citet{Gil09}; 
(37) \citet{Smi13}; 
(38) \citet{Max13}; 
(39) \citet{Sma12}; 
(40) \citet{Que10}. 
}
\end{landscape}
\end{longtab}

\subsection{Abundances}

Chemical abundances were determined for 12 refractory elements (Na, Mg, Al, Si, Ca, Ti, Cr, Ni, Co, Sc, Mn, and V), and lithium. For chromium, scandium, and titanium, we also calculated the abundance of the ions. The analysis for the refractory elements was again done in LTE, which is a good approximation for this stellar sample \citep{Ber12,Ser13}. We derived the abundances with the 2010 version of MOOG \citep{Sne73} and a grid of ATLAS plane-parallel model atmospheres \citep{Kur93}, using the EWs of the lines. For all elements these EWs were calculated with ARES. The final abundance for each element was calculated as the average value of the abundances given by each detected line of that element. The Li abundances, A(Li) = $\log$(N(Li)/N(H)) + 12, were derived by a standard LTE analysis using spectral synthesis with the revised version of the spectral synthesis code MOOG2010 \citep{Sne73}, a grid of Kurucz ATLAS9 atmospheres with overshooting \citep{Kur93}, and the linelist from \citet{Ghe09}. More details about these methods can be found in the works of \citet{Adi12} and \citet{Del13}.

All abundances can be found in Table \ref{TabAbu}. Several lithium abundances present upper limits since the lines are at the same level as the noise. The typical error for A(Li) is 0.1\,dex.

\begin{table*}
\caption{\label{TabAbu} Abundances for the transit hosts in this sample. }%This is only part of the table, as an example. The complete table can be found online.}
\centering
\begin{tabular}{cccccccccccc}

\hline\hline
Name & \ion{Al}{i} & \ion{Ca}{i} & \ion{Co}{i} & \ion{Cr}{i} & \ion{Cr}{ii} & \ion{Mg}{i} & \ion{Mn}{i} & $\ldots$ \\ % \ion{Na}{i} \\
 & (dex) & (dex) & (dex) & (dex) & (dex) & (dex) & (dex) & $\ldots$ \\ %  (dex) \\
\hline
CoRoT-9 & -0.03 $\pm$ 0.04 & 0.01 $\pm$ 0.06 & 0.00 $\pm$ 0.05 & 0.01 $\pm$ 0.06 & -0.08 $\pm$ 0.07 & -0.02 $\pm$ 0.02 & -0.04 $\pm$ 0.03 &  $\ldots$ \\ % -0.06 $\pm$ 0.08 \\
WASP-31 & -0.30 $\pm$ 0.22 & -0.06 $\pm$ 0.19 & 0.00 $\pm$ 0.19 & -0.01 $\pm$ 0.10 & -0.10 $\pm$ 0.12 & -0.13 $\pm$ 0.08 & -0.22 $\pm$ 0.14 & $\ldots$ \\ %  -0.28 $\pm$ 0.06 \\
$\ldots$\\

% \hline\hline
% Name & \ion{Ni}{i} & \ion{Sc}{i} & \ion{Sc}{ii} & \ion{Si}{i} & \ion{Ti}{i} & \ion{Ti}{ii} & \ion{V}{i} & \ion{Li}{i} \\
%  & (dex) & (dex) & (dex) & (dex) & (dex) & (dex) & (dex) & (dex) \\
% \hline
% CoRoT-9 & -0.04 $\pm$ 0.05 & 0.05 $\pm$ 0.02 & -0.08 $\pm$ 0.06 & -0.01 $\pm$ 0.05 & 0.04 $\pm$ 0.04 & -0.05 $\pm$ 0.04 & 0.04 $\pm$ 0.04 & $\leq$ 0.88 \\
% WASP-31 & -0.17 $\pm$ 0.15 & 0.21 $\pm$ 0.03 & -0.02 $\pm$ 0.11 & -0.09 $\pm$ 0.11 & -0.03 $\pm$ 0.16 & 0.16 $\pm$ 0.04 & -0.04 $\pm$ 0.27 & 3.02 \\
% $\ldots$\\

\hline\hline
Name & \ion{Al}{i}$_{LC}$ & \ion{Ca}{i}$_{LC}$ & \ion{Co}{i}$_{LC}$ & \ion{Cr}{i}$_{LC}$ & \ion{Cr}{ii}$_{LC}$ & \ion{Mg}{i}$_{LC}$ & \ion{Mn}{i}$_{LC}$ &  $\ldots$ \\ % \ion{Na}{i}$_{LC}$ \\
 & (dex) & (dex) & (dex) & (dex) & (dex) & (dex) & (dex) &  $\ldots$ \\ % (dex) \\
\hline
CoRoT-9 & -0.03 $\pm$ 0.04 & -0.03 $\pm$ 0.06 & 0.01 $\pm$ 0.05 & 0.00 $\pm$ 0.07 & -0.04 $\pm$ 0.07 & -0.04 $\pm$ 0.03 & -0.05 $\pm$ 0.03 & $\ldots$ \\ %  -0.07 $\pm$ 0.07 \\
WASP-31 & -0.29 $\pm$ 0.22 & -0.02 $\pm$ 0.20 & 0.00 $\pm$ 0.19 & 0.00 $\pm$ 0.10 & -0.26 $\pm$ 0.13 & -0.10 $\pm$ 0.11 & -0.21 $\pm$ 0.14 & $\ldots$ \\ %  -0.27 $\pm$ 0.06 \\
$\ldots$\\

% \hline\hline
% Name & \ion{Ni}{i}$_{LC}$ & \ion{Sc}{i}$_{LC}$ & \ion{Sc}{ii}$_{LC}$ & \ion{Si}{i}$_{LC}$ & \ion{Ti}{i}$_{LC}$ & \ion{Ti}{ii}$_{LC}$ & \ion{V}{i}$_{LC}$ & \ion{Li}{i}$_{LC}$ \\
%  & (dex) & (dex) & (dex) & (dex) & (dex) & (dex) & (dex) & (dex) \\
% \hline
% CoRoT-9 & -0.03 $\pm$ 0.05 & 0.05 $\pm$ 0.02 & -0.04 $\pm$ 0.06 & 0.00 $\pm$ 0.05 & 0.04 $\pm$ 0.04 & 0.01 $\pm$ 0.03 & 0.04 $\pm$ 0.04 & $\leq$ 0.87 \\
% WASP-31 & -0.17 $\pm$ 0.15 & 0.22 $\pm$ 0.03 & -0.18 $\pm$ 0.11 & -0.08 $\pm$ 0.11 & -0.02 $\pm$ 0.16 & -0.01 $\pm$ 0.04 & -0.03 $\pm$ 0.27 & 3.03 \\
% $\ldots$\\
\hline
\end{tabular}
\tablefoot{The complete table is provided in electronic form at the CDS.}
\end{table*}

\subsection{Masses and radii}

Stellar masses and radii were computed with the calibration of \citet{Tor10}. This calibration is based on effective temperature, surface gravity and metallicity. For stellar mass, we applied a small quadratic correction. \citet{Tor10} and \citet{San13} show that there is a small offset between masses obtained through this calibration and masses obtained through isochrones. \citet{San13} fit this offset with a quadratic function that we use to correct the masses obtained through the calibration of \citet{Tor10}:

\begin{equation}
M_{cor} = 0.791 \cdot M_T^2 - 0.575 \cdot M_T + 0.701
\end{equation}

\noindent where $M_{cor}$ and $M_T$ denote the corrected stellar masses and the mass from the \citet{Tor10} calibration, respectively. %We also computed the radii through Newton's law of gravitation.
Table \ref{TabPar} lists all stellar parameters for the stars in this sample.

\section{Photometric surface gravity}\label{logg}

\begin{figure}[t!]
\begin{center}
\includegraphics[width=7.0cm]{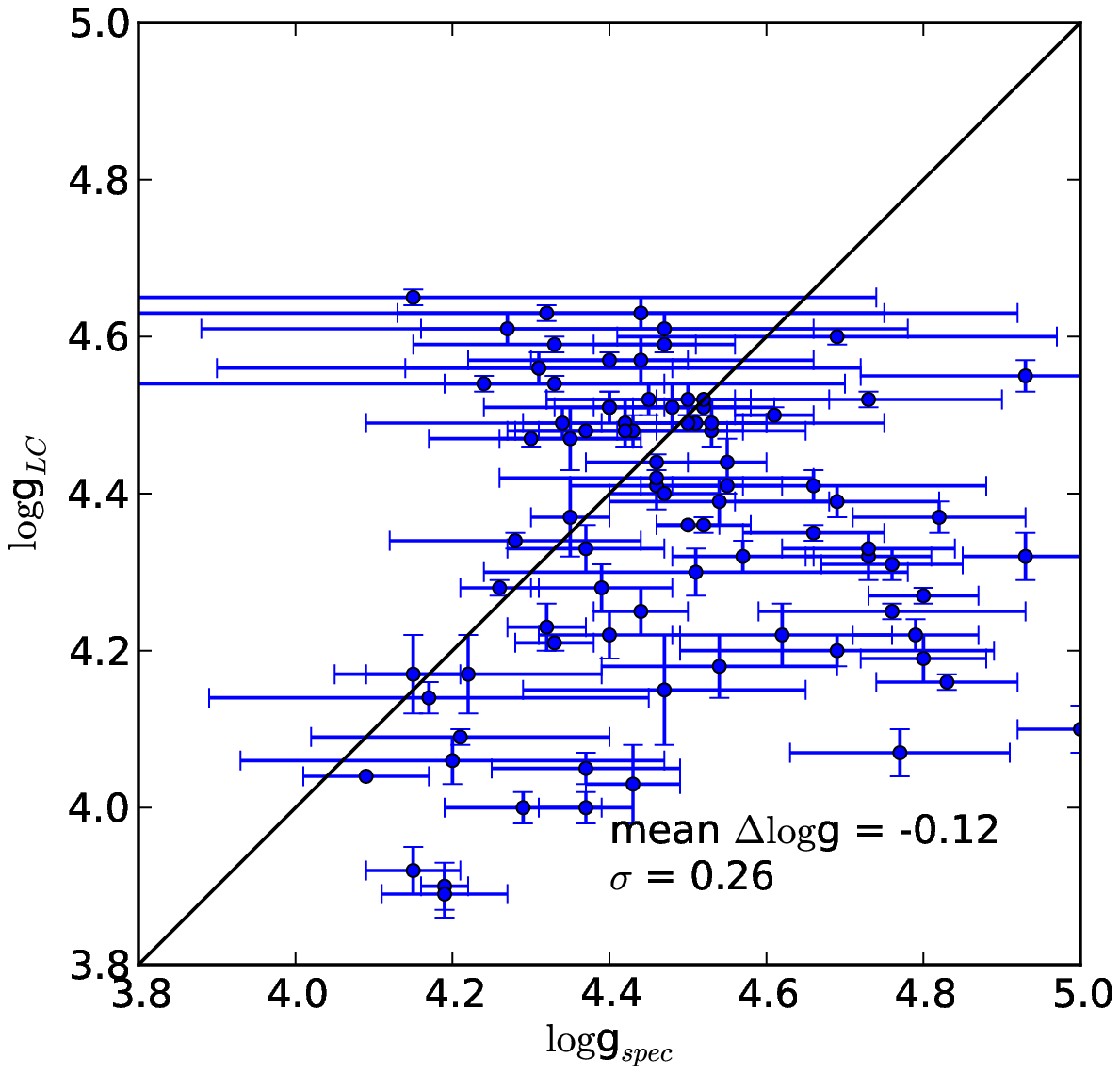}
\includegraphics[width=7.0cm]{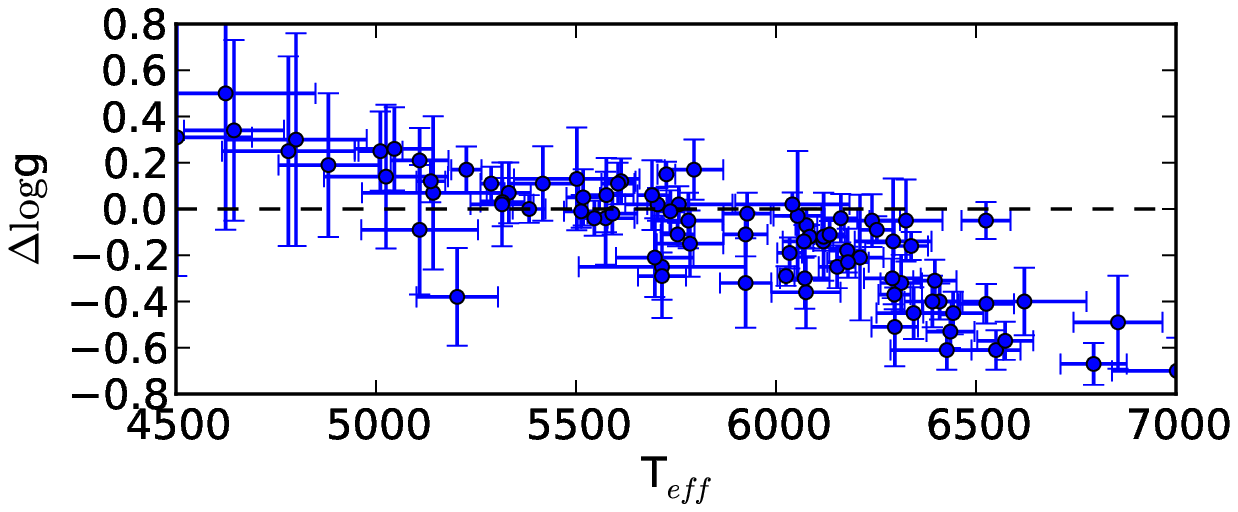}\caption{Top panel: Comparison of the spectroscopic and the photometric surface gravity. Bottom panel: Differences in $\log$g (defined as `photometric - spectroscopic') as a function of the effective temperature.}
\label{FigLogg}
\end{center}
\end{figure}

Over the years, it has become clear that determining surface gravities spectroscopically is not well constrained \citep[e.g. ][]{Soz07,Tor12}. Luckily, for stars with a transiting planet, the photometric light curve can be used independently to determine the surface gravity with much better precision. This can improve the precision of the stellar mass and radius and consequently also the precision of the planetary mass and radius. Good precision is necessary for a correct classification of the exoplanets. Purely from transit photometry, the stellar density can be calculated from Kepler's third law \citep{Sea03}:

\begin{equation}
\rho_{\ast} + k^3\rho_p = \frac{3\pi}{GP^2}\left(\frac{a}{R_{\ast}}\right)^3
\end{equation}

\noindent where $\rho_{\ast}$ and $\rho_p$ are the stellar and planetary density, $P$ the period of the planet, $a$ the orbital separation, $G$ the gravitational constant, and $R_{\ast}$ the stellar radius. Since the constant coefficient $k$ is usually small, the second term on the left is negligible. All parameters on the right come directly from analyzing the transit light curve.

With this stellar density, combined with the effective temperature and metallicity from the spectroscopic analysis, the surface gravity can be determined through isochrone fitting, as described in \citet{Soz07}. For this work, we used the stellar densities from the discovery papers, PARSEC isochrones \citep{Bre12}, a $\chi^2$ minimization process for the fitting, and the individual metallicity and effective temperature from our spectroscopic analysis. The one-sigma error bars were computed using all solutions where $\chi^2 < 3$. From all these solutions, we computed the standard deviation of all surface gravities.

All values can be found in Table \ref{TabPar}. For WASP-45, WASP-56, and XO-2, no photometric surface gravity could be calculated owing to the high metallicity and the uncertainties of the models at these high metallicities \citep{Val13}. In the top panel of Figure \ref{FigLogg}, we compare the spectroscopic and the photometric surface gravity. It can be seen that they do not always compare well. The differences in surface gravity also depend on the temperature as can be seen in the bottom panel of Figure \ref{FigLogg}, where a decreasing linear trend is noticeable. The same trend is found for the microturbulence, which is closely related to the temperature. Comparing the $\log$g differences with metallicities reveals no additional trends. These trends and the possible causes will be discussed in a forthcoming work.

Photometric surface gravities are generally more precise than spectroscopic surface gravities. This higher precision, however, does not guarantee higher accuracy. To determine the stellar density, which is used to derive the photometric surface gravity, the ratio $a/R_{\ast}$ is used. This value comes from fitting the light curve, which depends on a correct limb darkening coefficient. This limb darkening coefficient can be fixed using the dependence on the effective temperature. An incorrect effective temperature will thus lead to an incorrect fixed limb darkening coefficient and thus an incorrect fitting of the light curve. However, the limb darkening coefficient can also be left as a free parameter in the fit. The determination of $a/R_{\ast}$ also depends on the orbital eccentricity. This eccentricity is determined from a radial velocity curve and is thus unfortunately not always known for transiting planets and fixed to a standard value in the transit light curve fit. Furthermore, the photometric surface gravities depend on theoretical stellar evolution models. The spectroscopic surface gravities are poorly constrained and thus not necessarily accurate either. Since both methods have their pros and cons, we provide the reader with both values.

The other atmospheric parameters that are spectroscopically determined are much better constrained, so we adopt these parameters for the continuation of this work. Effective temperatures derived with our method, have shown to compare well with well established methods, such as the IRFM \citep[e.g. ][]{Tsa13,San13}. \citet{Tor12} explored the impact of using different surface gravities on the other atmospheric parameters. They show that by using the method that we use in this work, the impact is minimal, compared with other methods. However, small trends are still present. These trends and their possible corrections will be explored in a forthcoming paper.

\subsection{Chemical abundances}

\begin{figure*}[t!]
\begin{center}
\includegraphics[width=4.5cm]{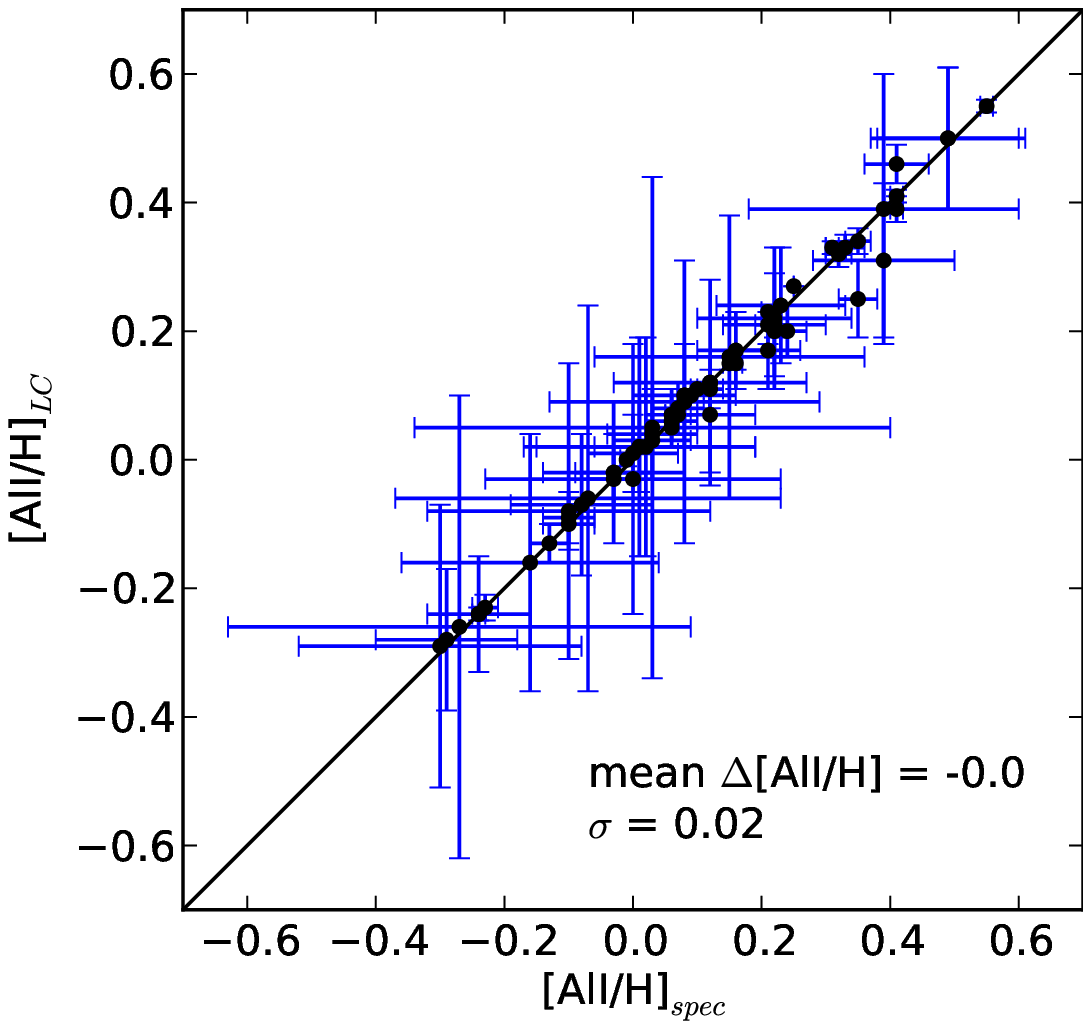}
\includegraphics[width=4.5cm]{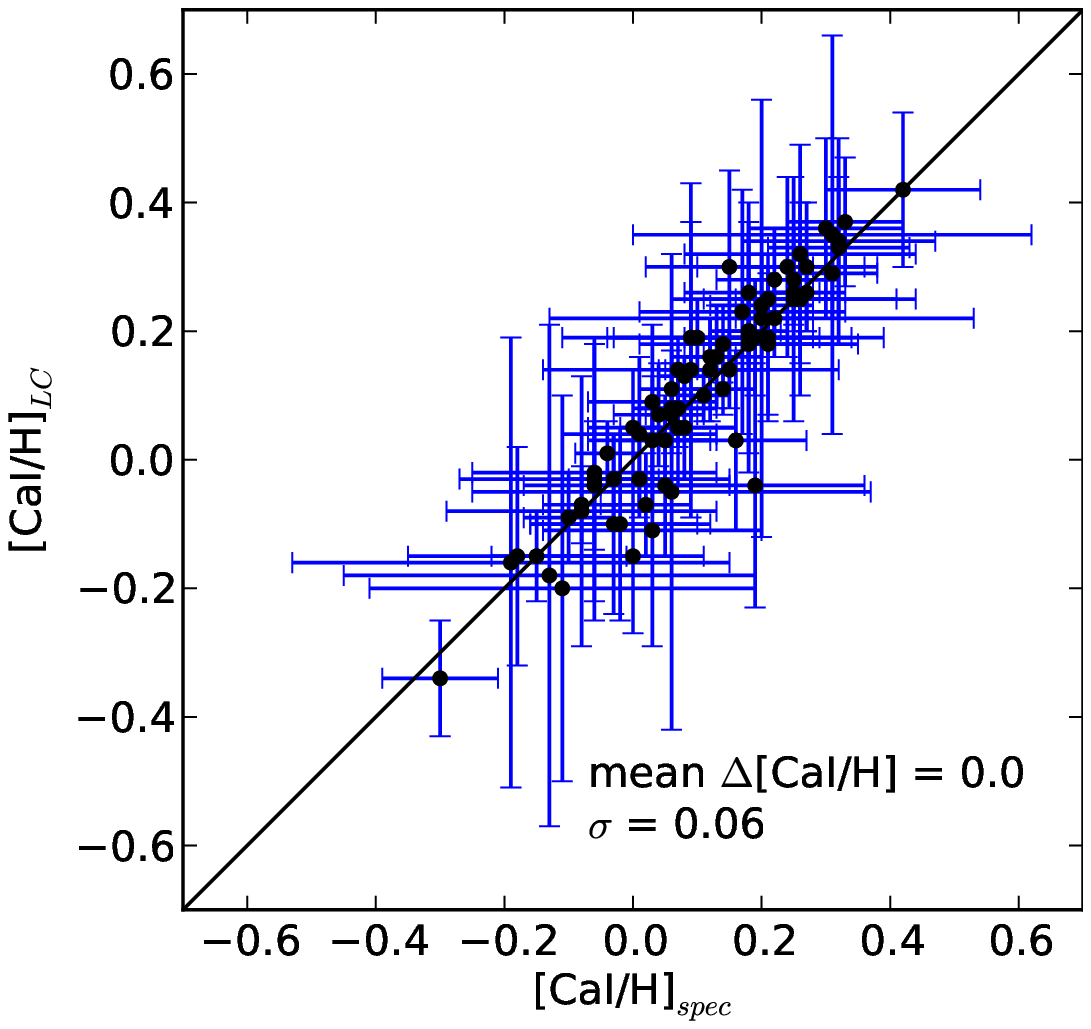}
\includegraphics[width=4.5cm]{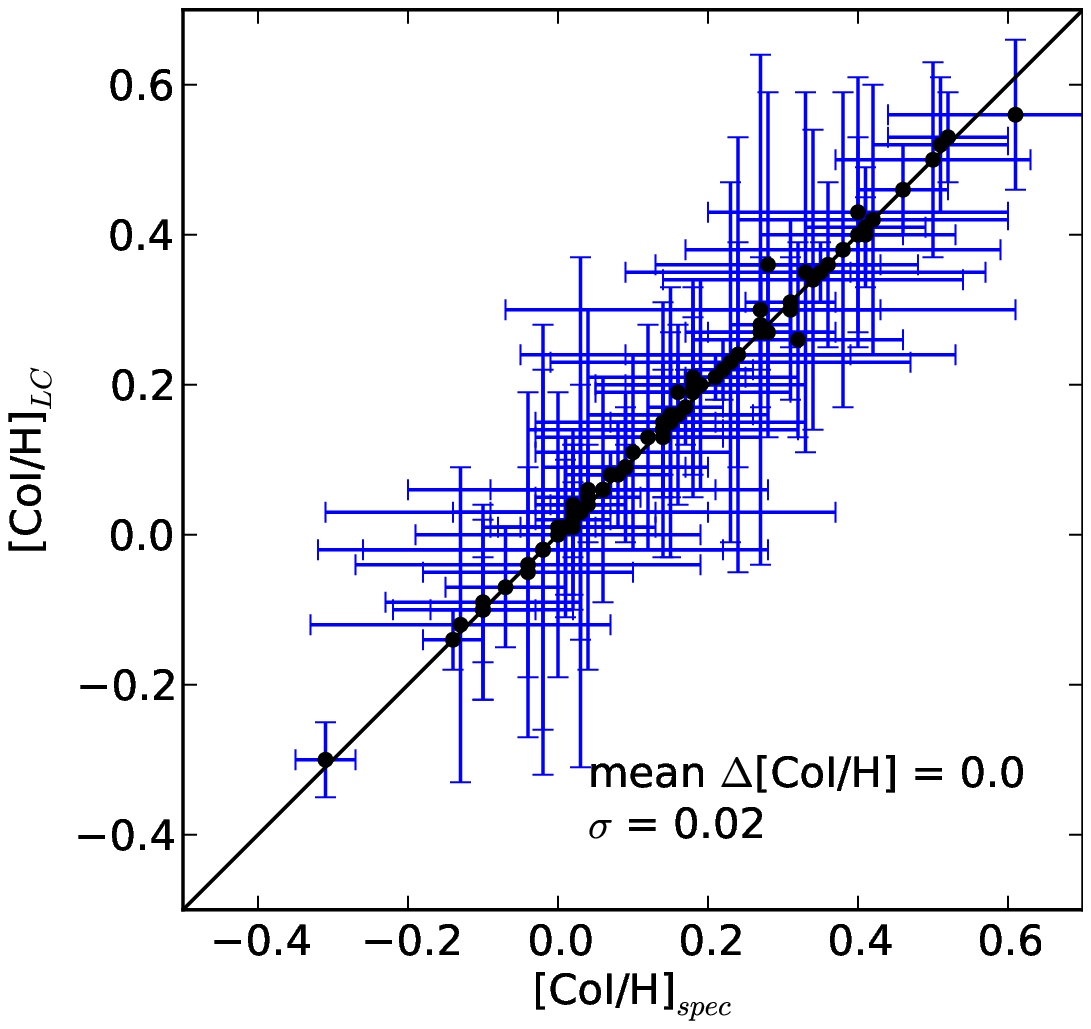}
\includegraphics[width=4.5cm]{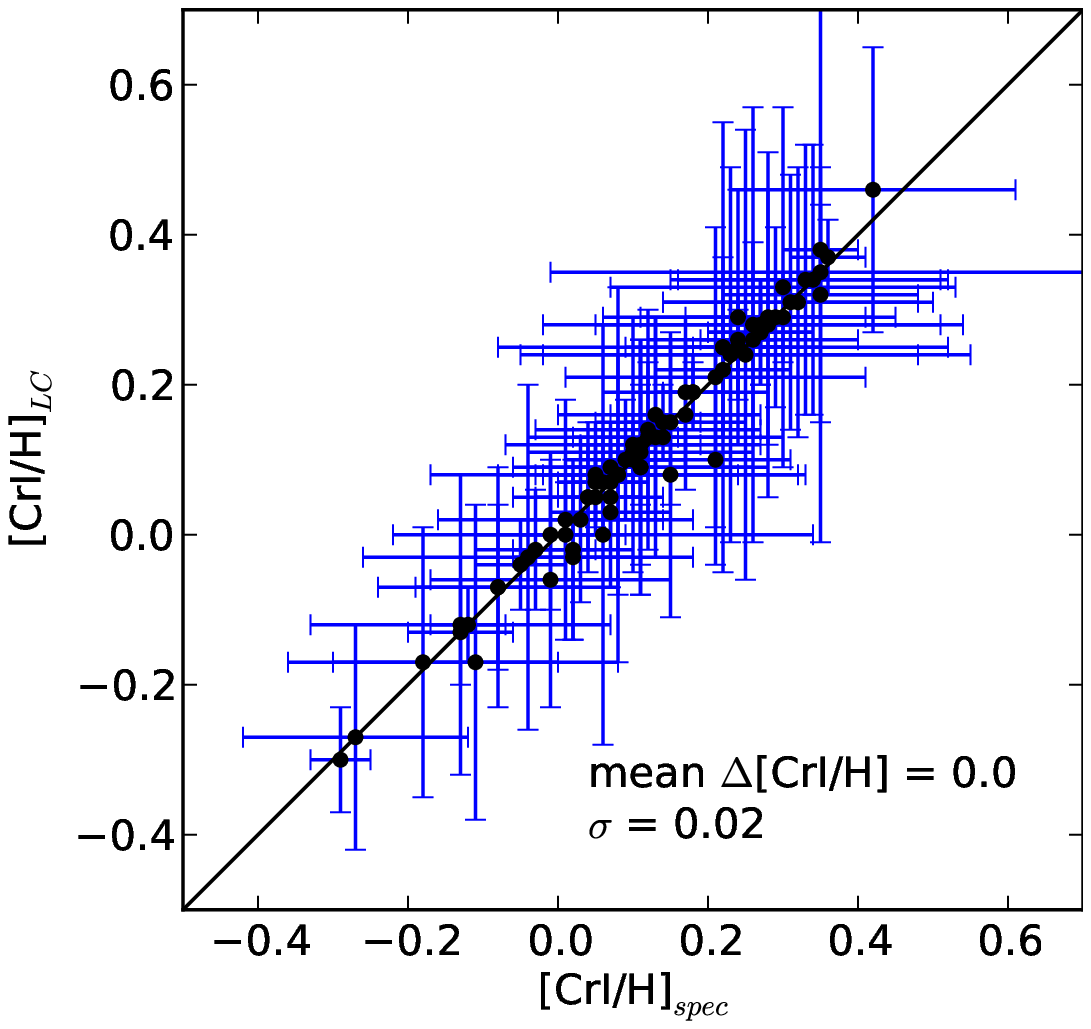}\\
\includegraphics[width=4.5cm]{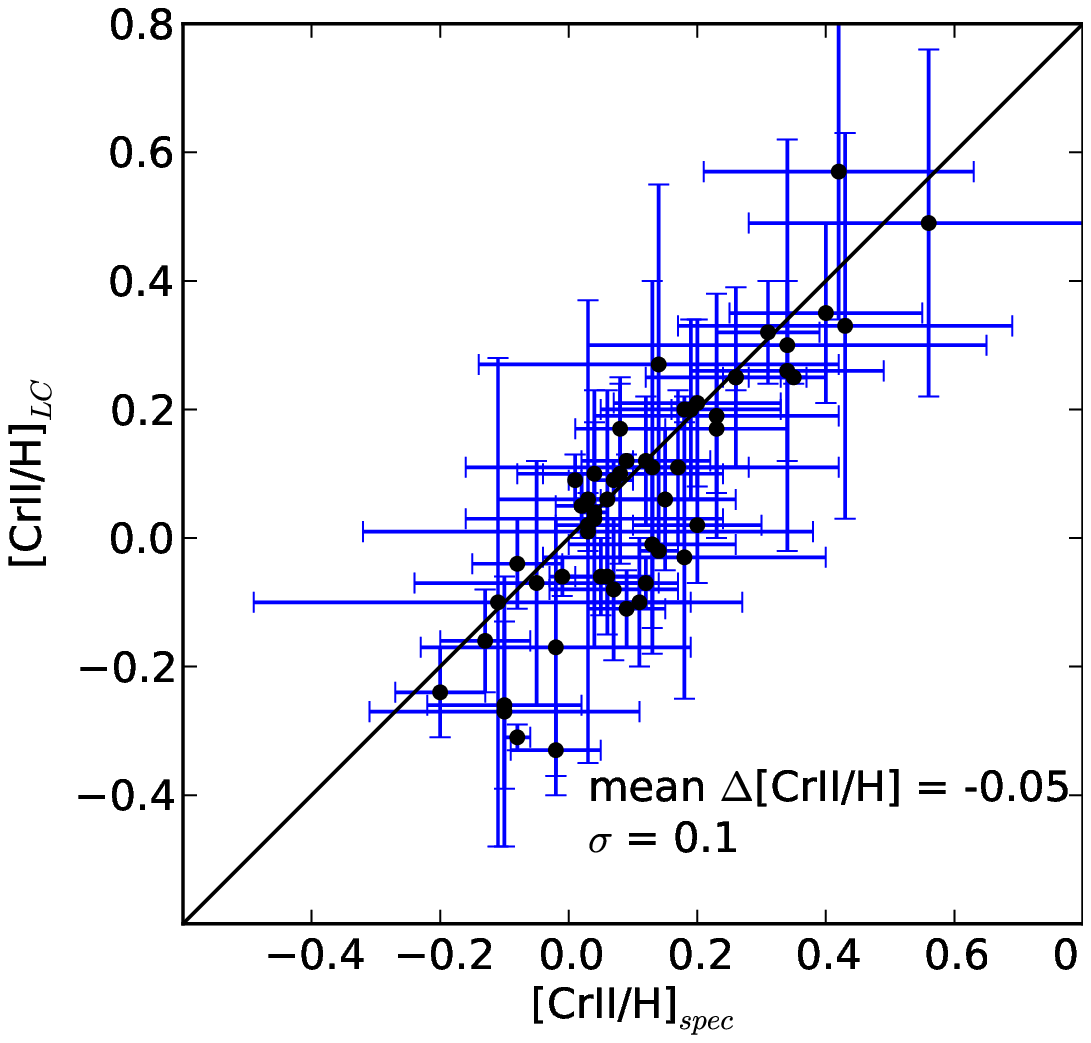}
\includegraphics[width=4.5cm]{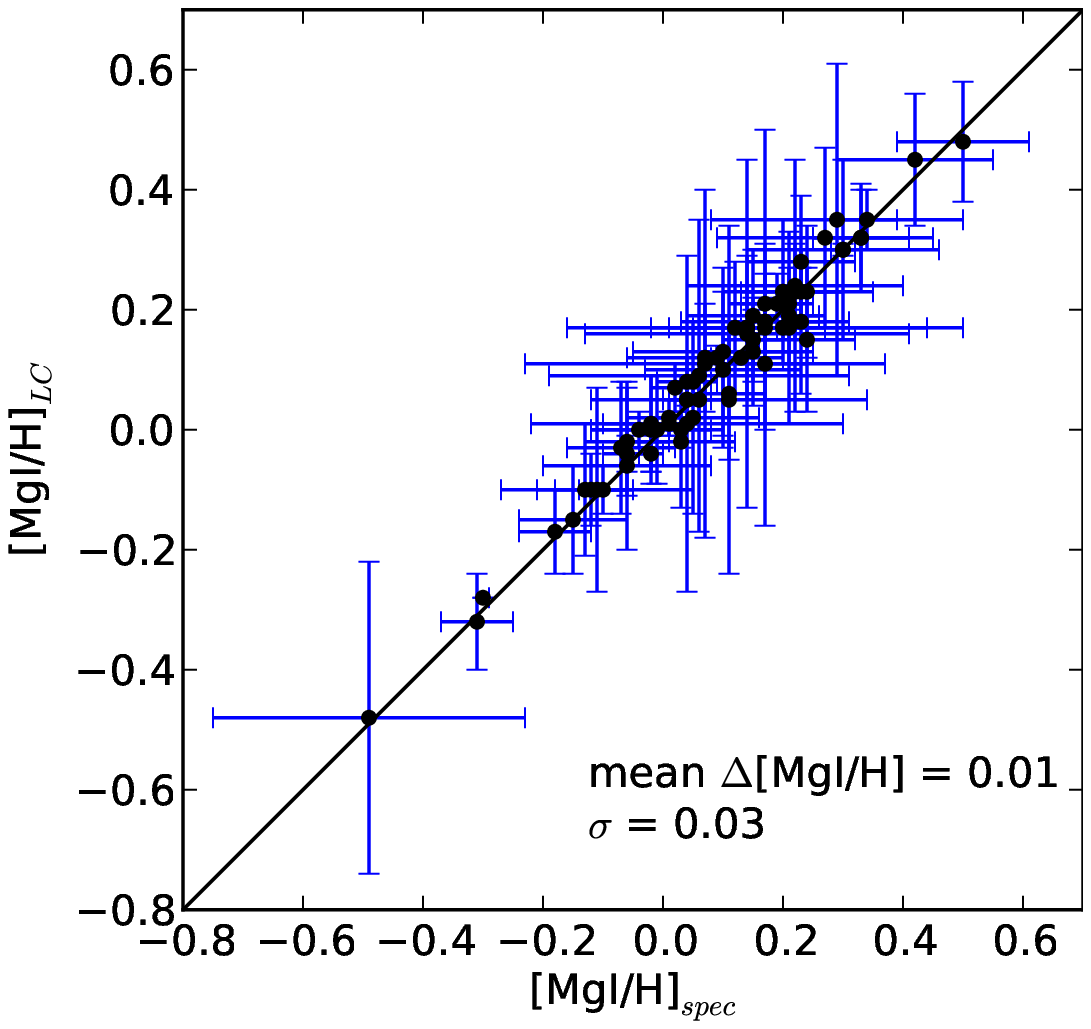}
\includegraphics[width=4.5cm]{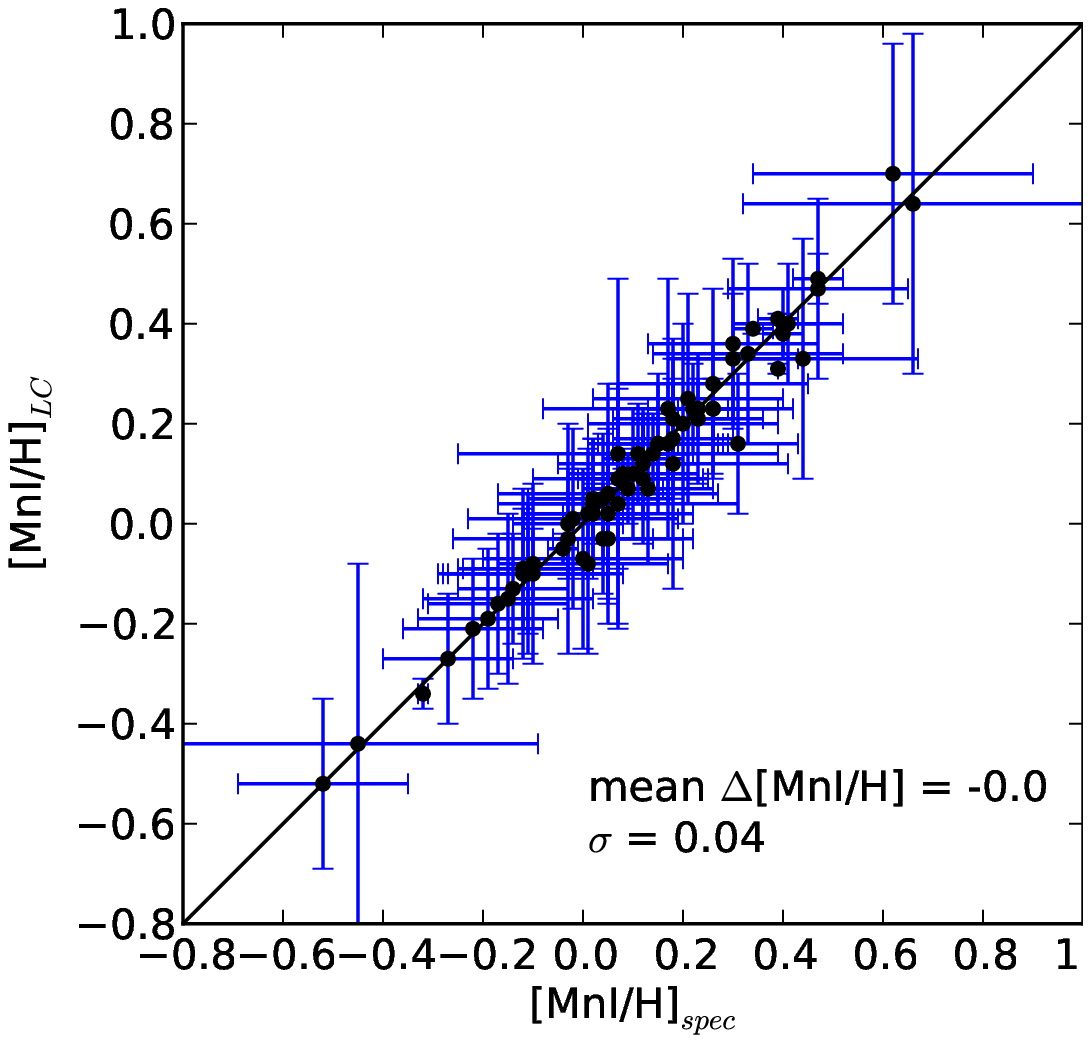}
\includegraphics[width=4.5cm]{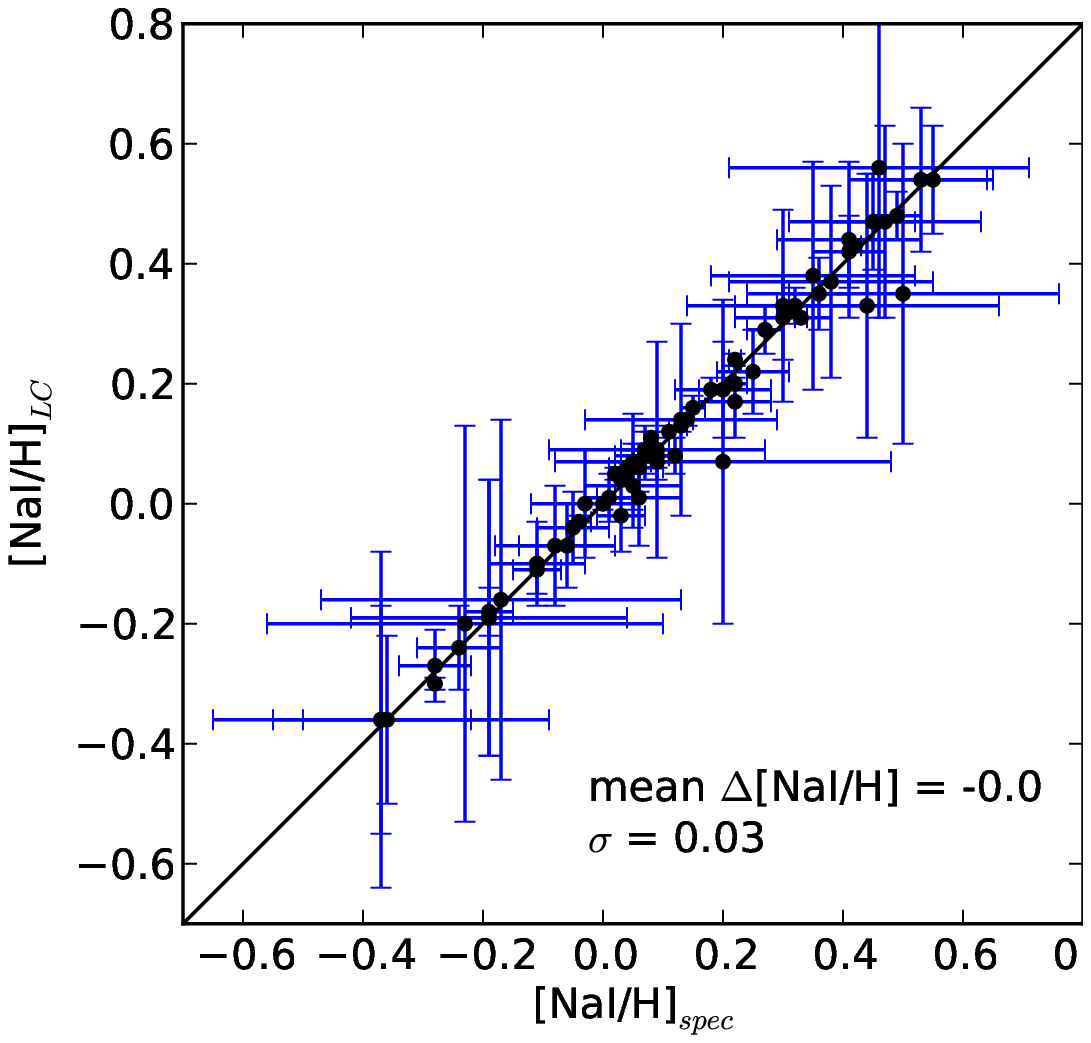}\\
\includegraphics[width=4.5cm]{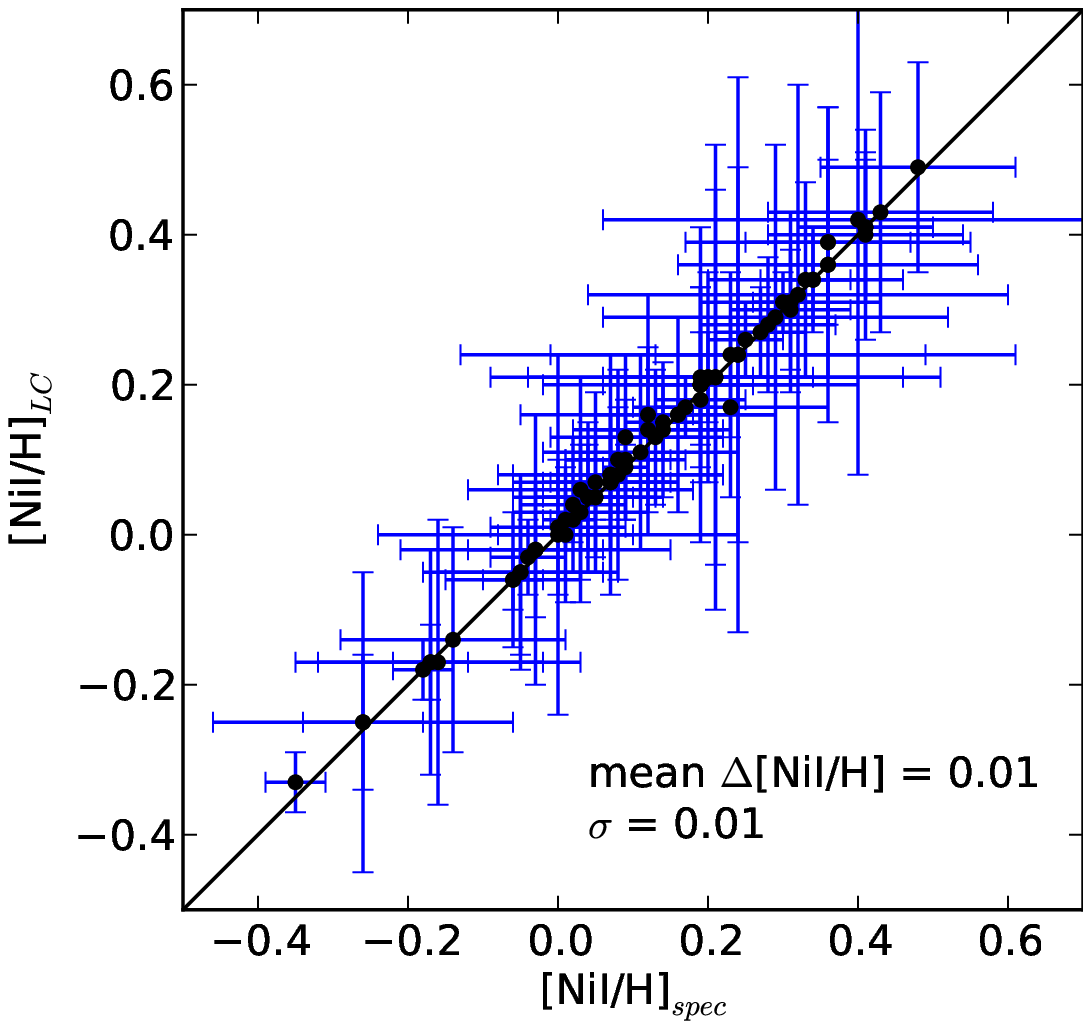}
\includegraphics[width=4.5cm]{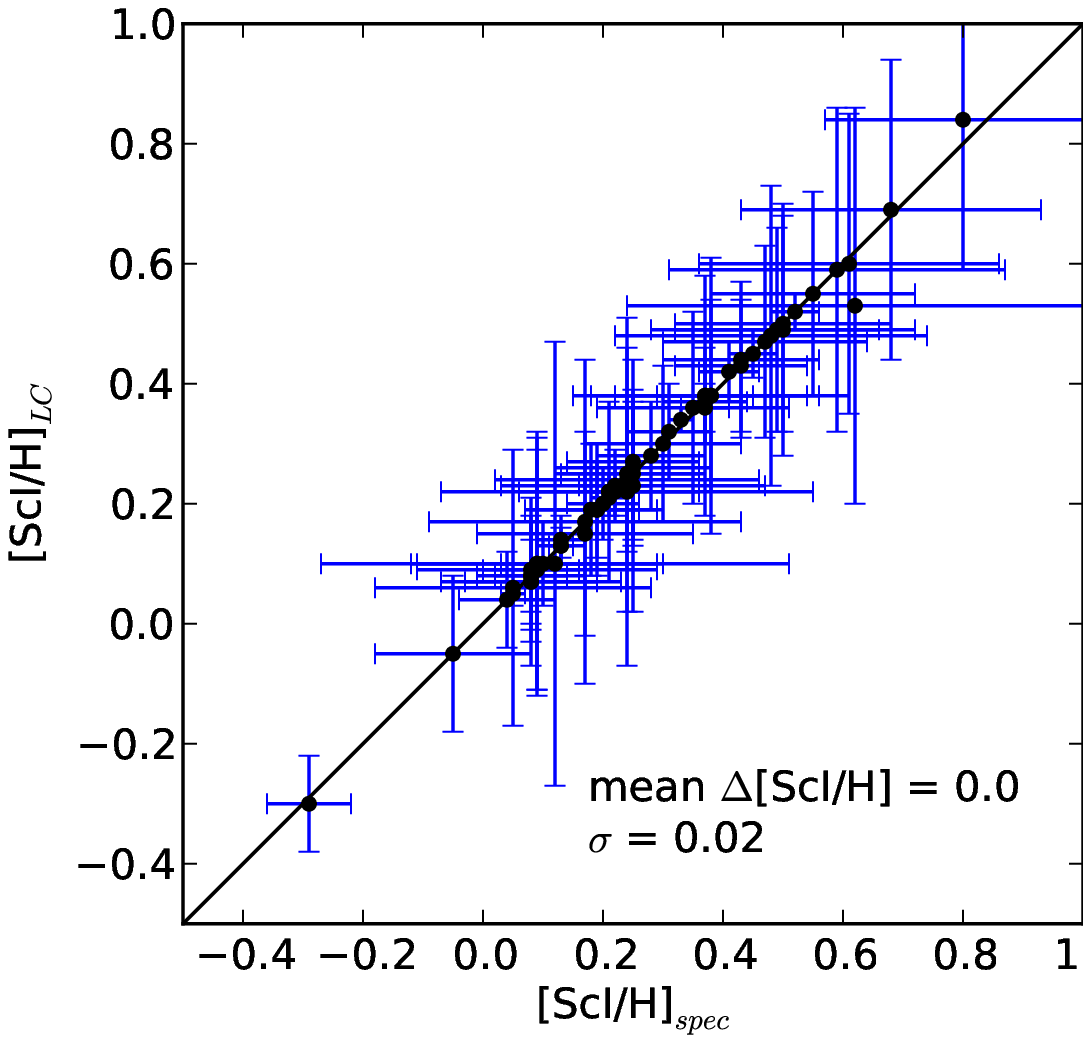}
\includegraphics[width=4.5cm]{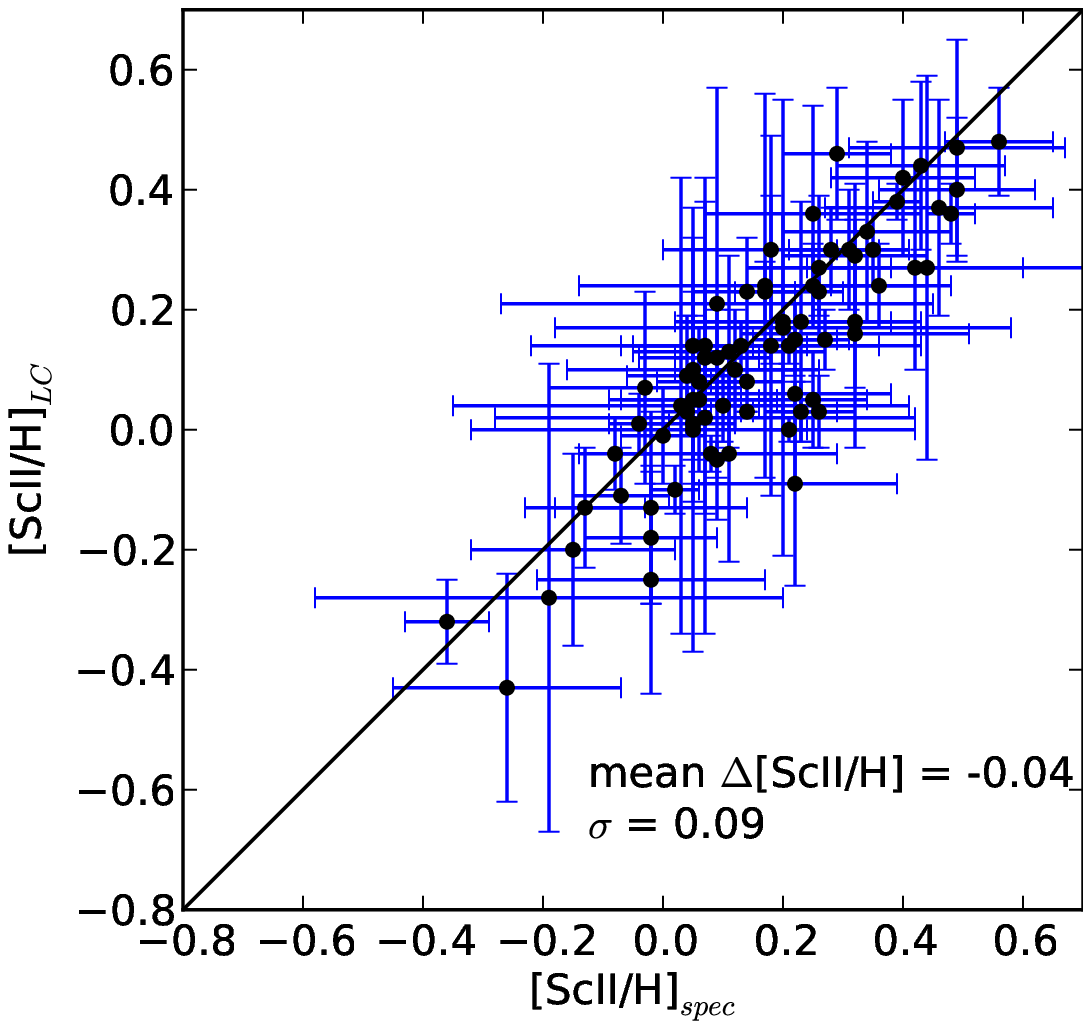}
\includegraphics[width=4.5cm]{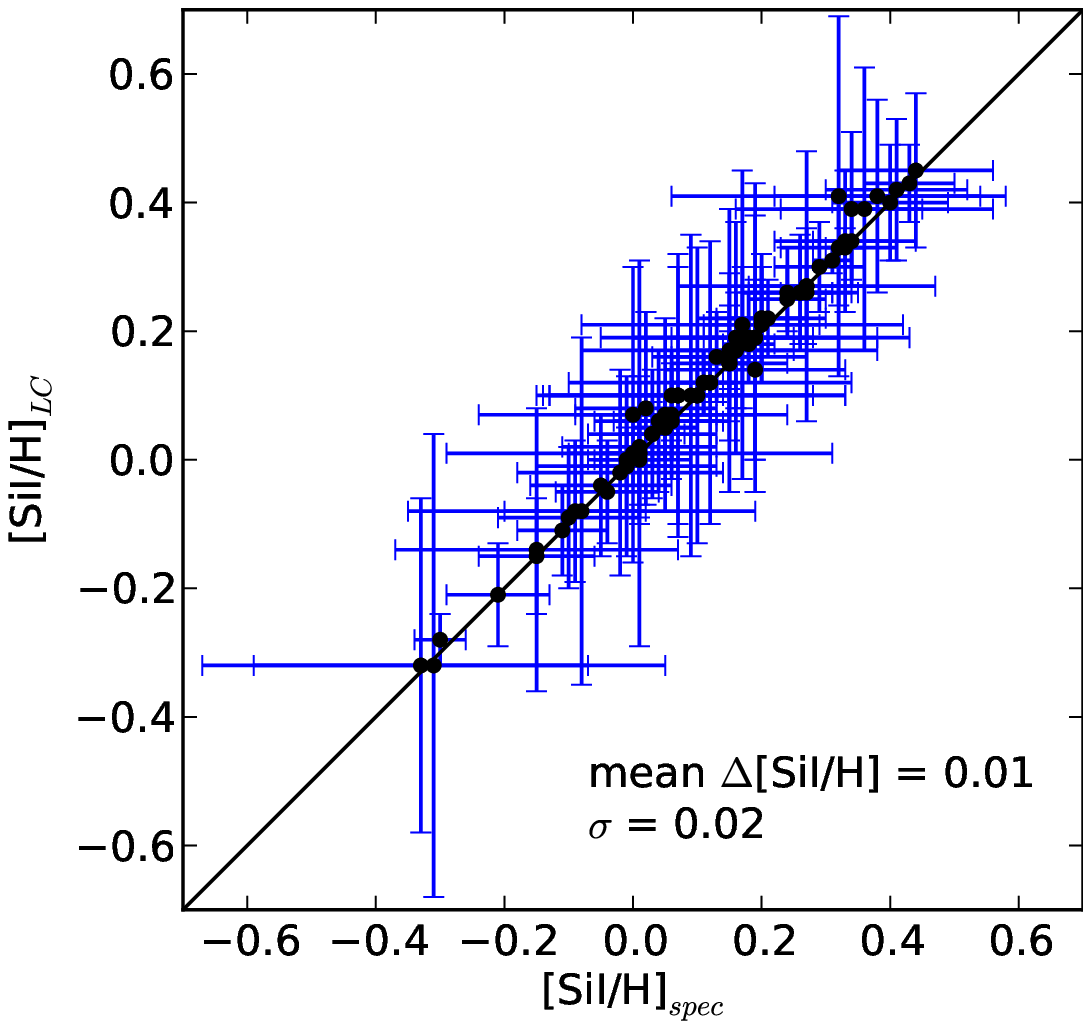}\\
\includegraphics[width=4.5cm]{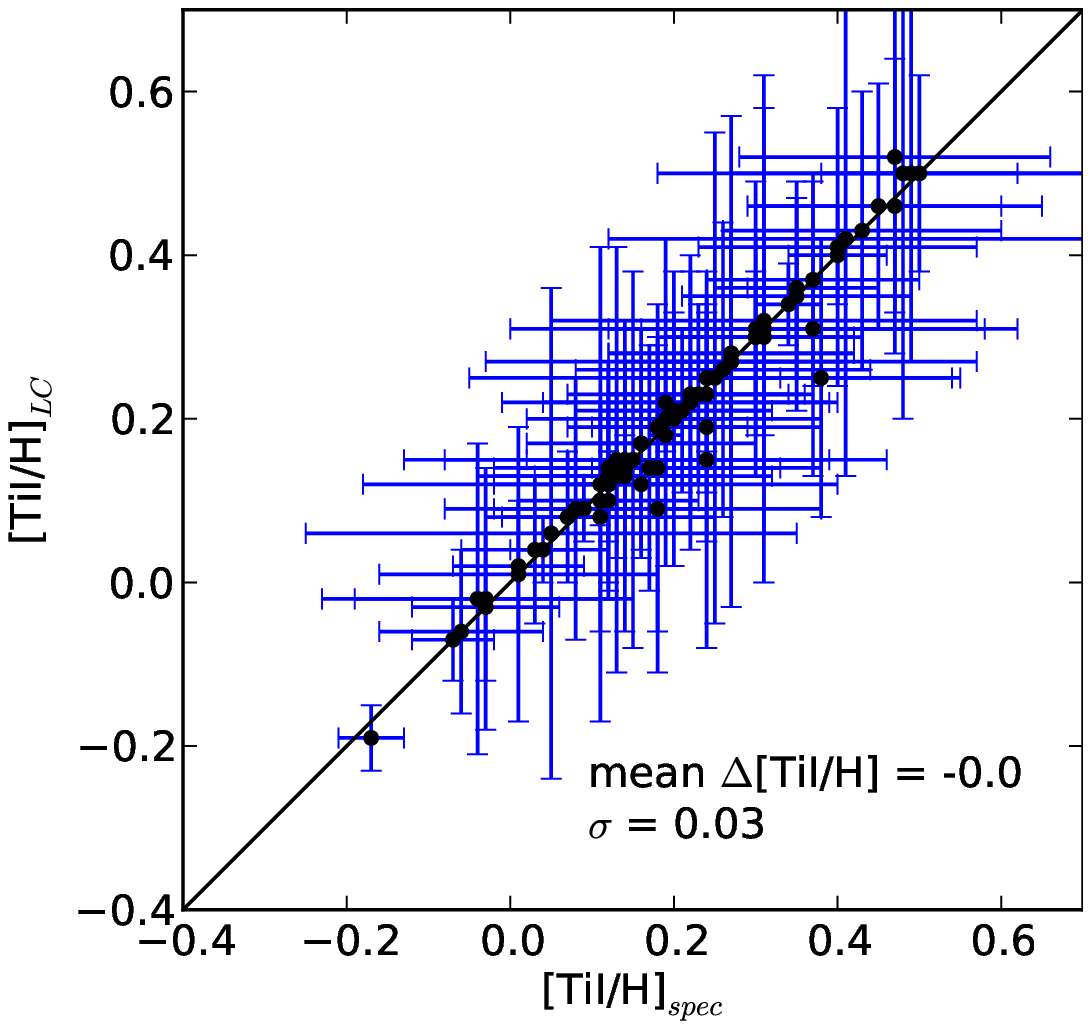}
\includegraphics[width=4.5cm]{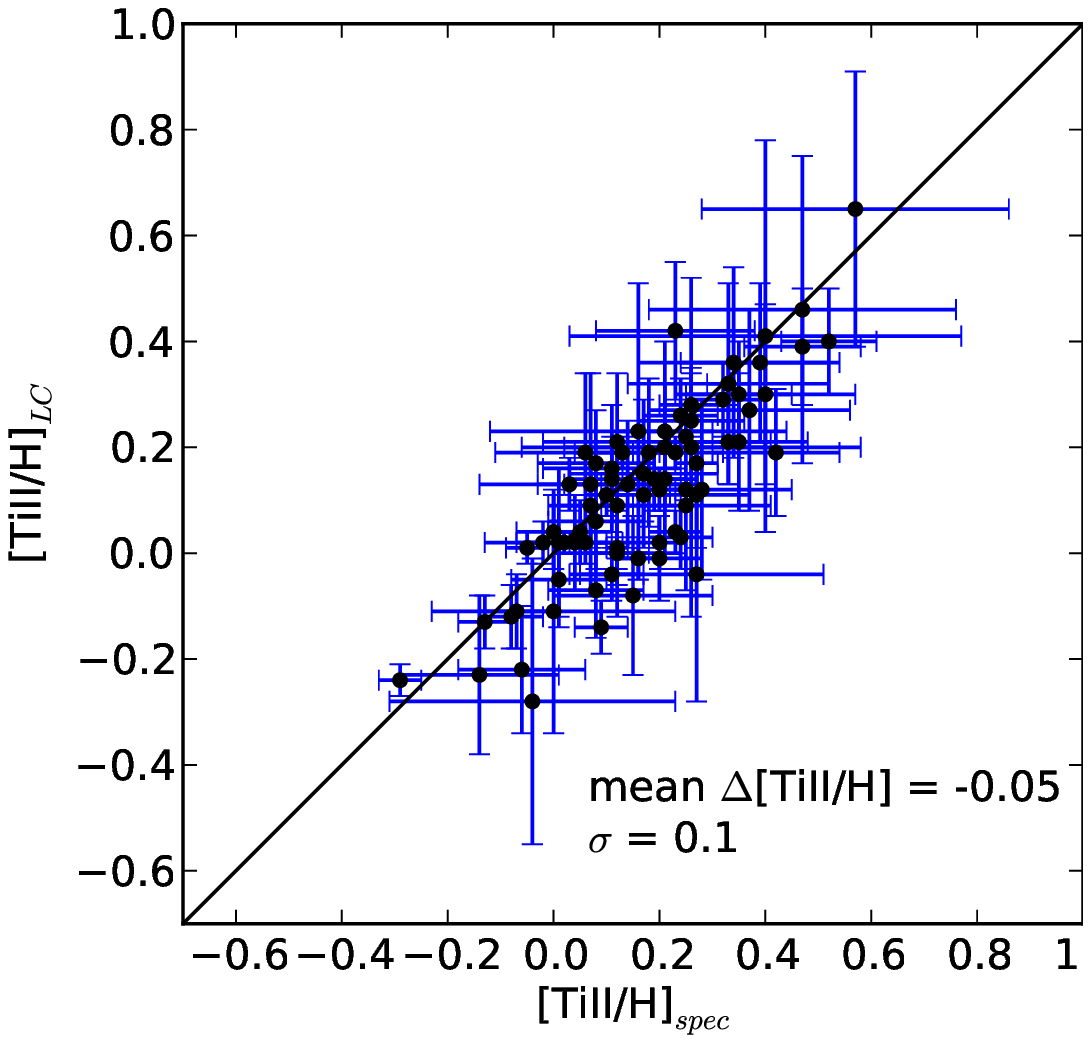}
\includegraphics[width=4.5cm]{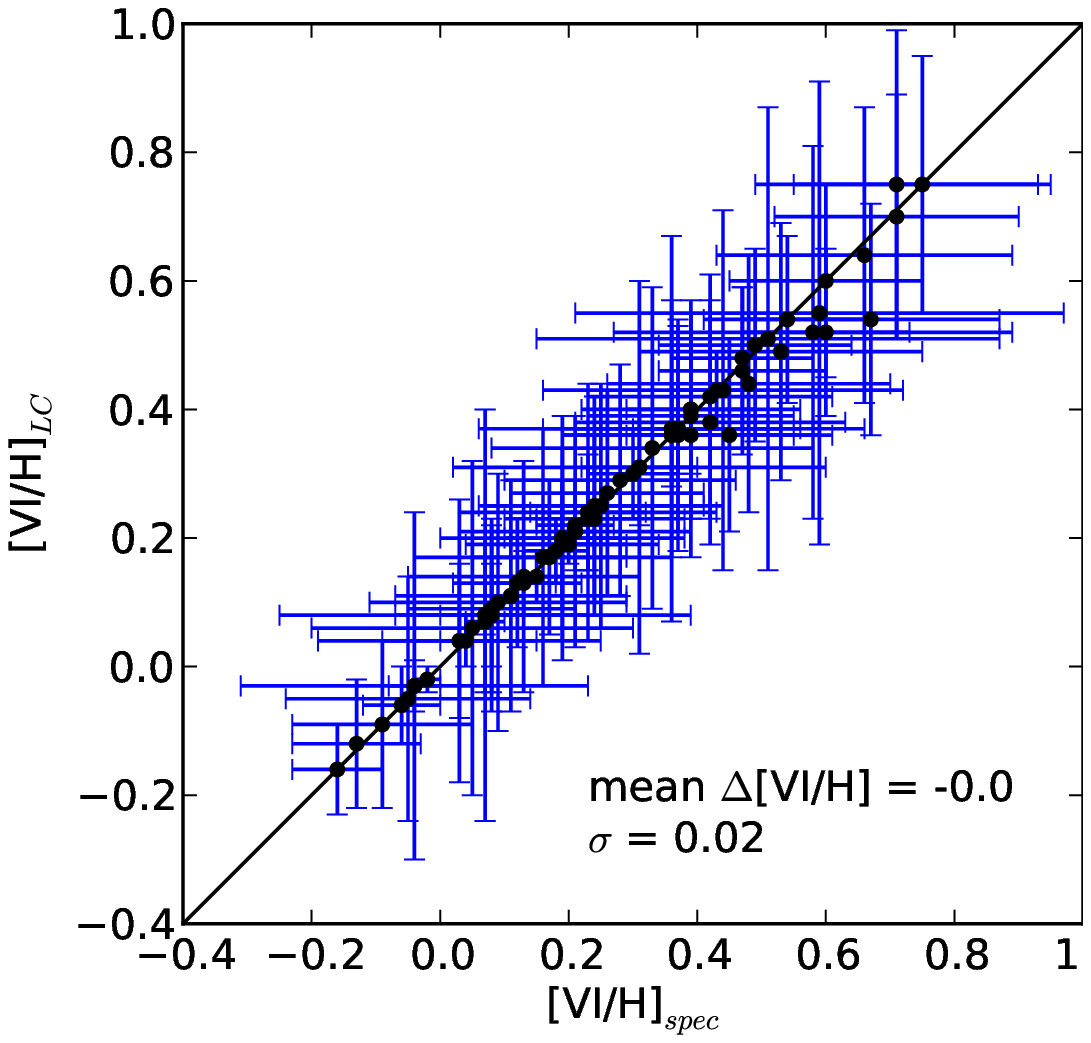}
\caption{Comparisons of the chemical abundances, obtained with the spectroscopic and photometric surface gravity. }
\label{FigAbu}
\end{center}
\end{figure*}

The derivation of the chemical abundances is based on all atmospheric parameters and thus also depends on the surface gravity. We recalculated the abundances {of the refractory elements} (see Table \ref{TabAbu}) with the photometric surface gravity. For all atomic elements, there is virtually no difference between the two abundances, as can be seen in Figure \ref{FigAbu}. Since the atom abundances do not differ, we did not redo the spectral synthesis to derive the lithium abundances $A(Li)$ again. For the three ions, on the other hand, the differences are greater. However, they are still within the error bars. Since ions are more sensitive to the surface gravity \citep{Gray92}, these larger differences are as expected.

\begin{figure}[t!]
\begin{center}
\includegraphics[width=6.5cm]{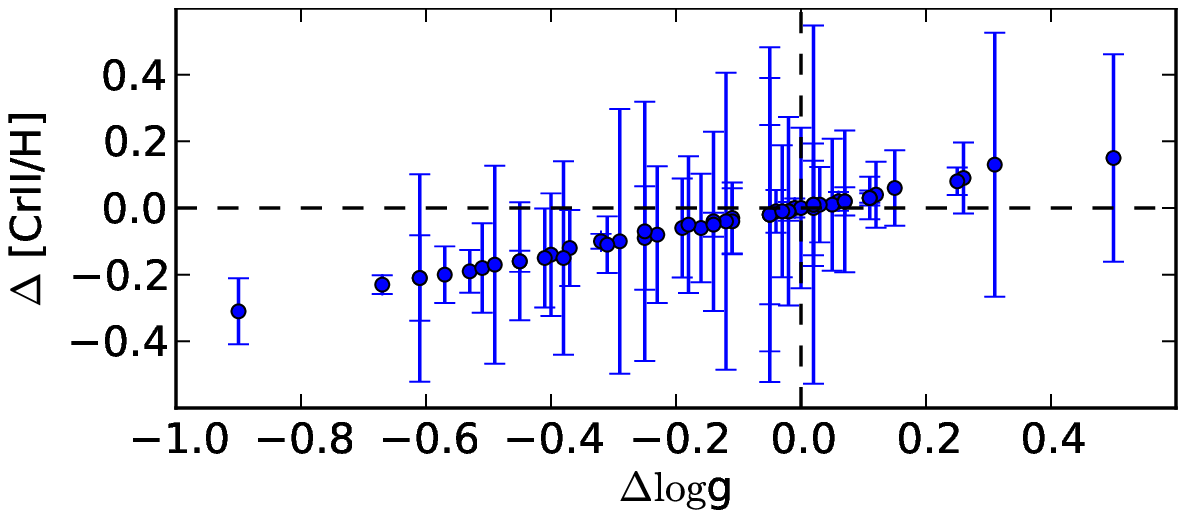}\\
\includegraphics[width=6.5cm]{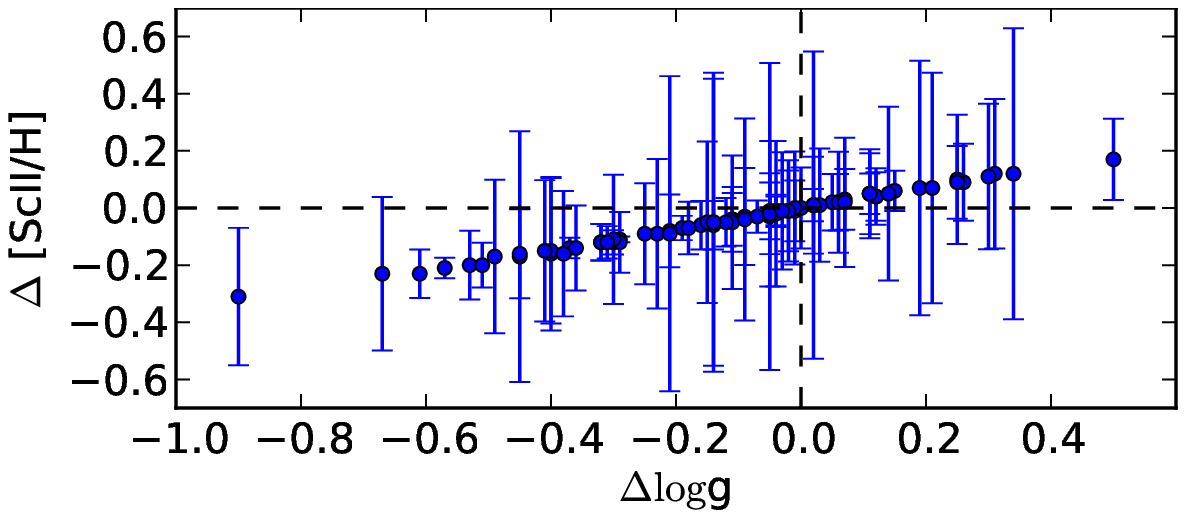}\\
\includegraphics[width=6.5cm]{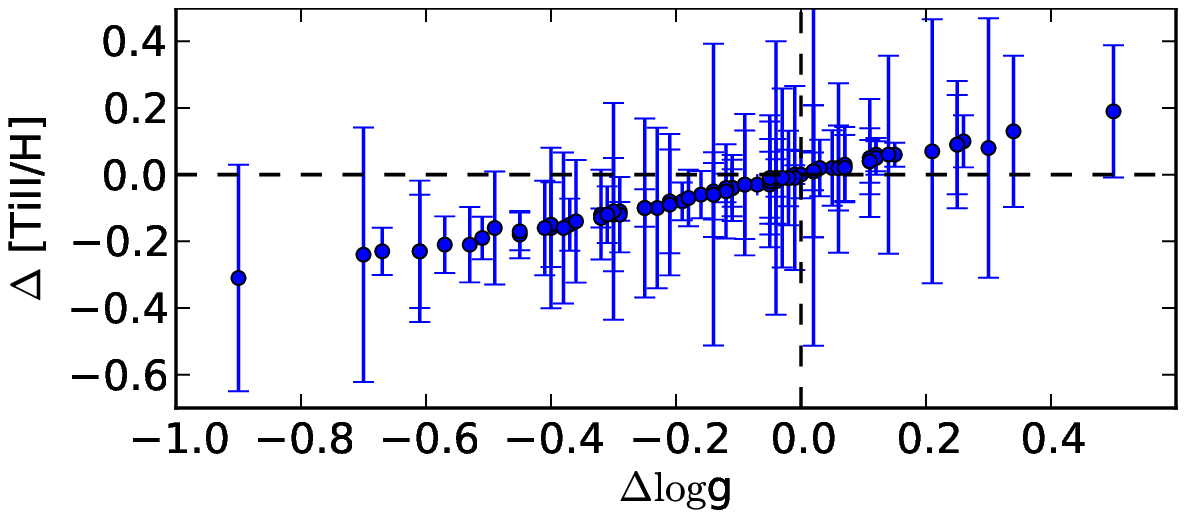}
\caption{Differences between the ion abundances as a function of the difference in $\log$g (defined as `constrained - unconstrained').}
\label{FigAbu2}
\end{center}
\end{figure}

Figure \ref{FigAbu2} shows the differences between these ion abundances as a function of the surface gravity difference. There are clear, visible linear trends with small slopes of $0.34$, $0.37$, and $0.37$ for \ion{Cr}{ii}, \ion{Sc}{ii}, and \ion{Ti}{ii}, respectively. 

\subsection{Stellar mass and radius}

\begin{figure*}[t!]
\begin{center}
\includegraphics[width=5.8cm]{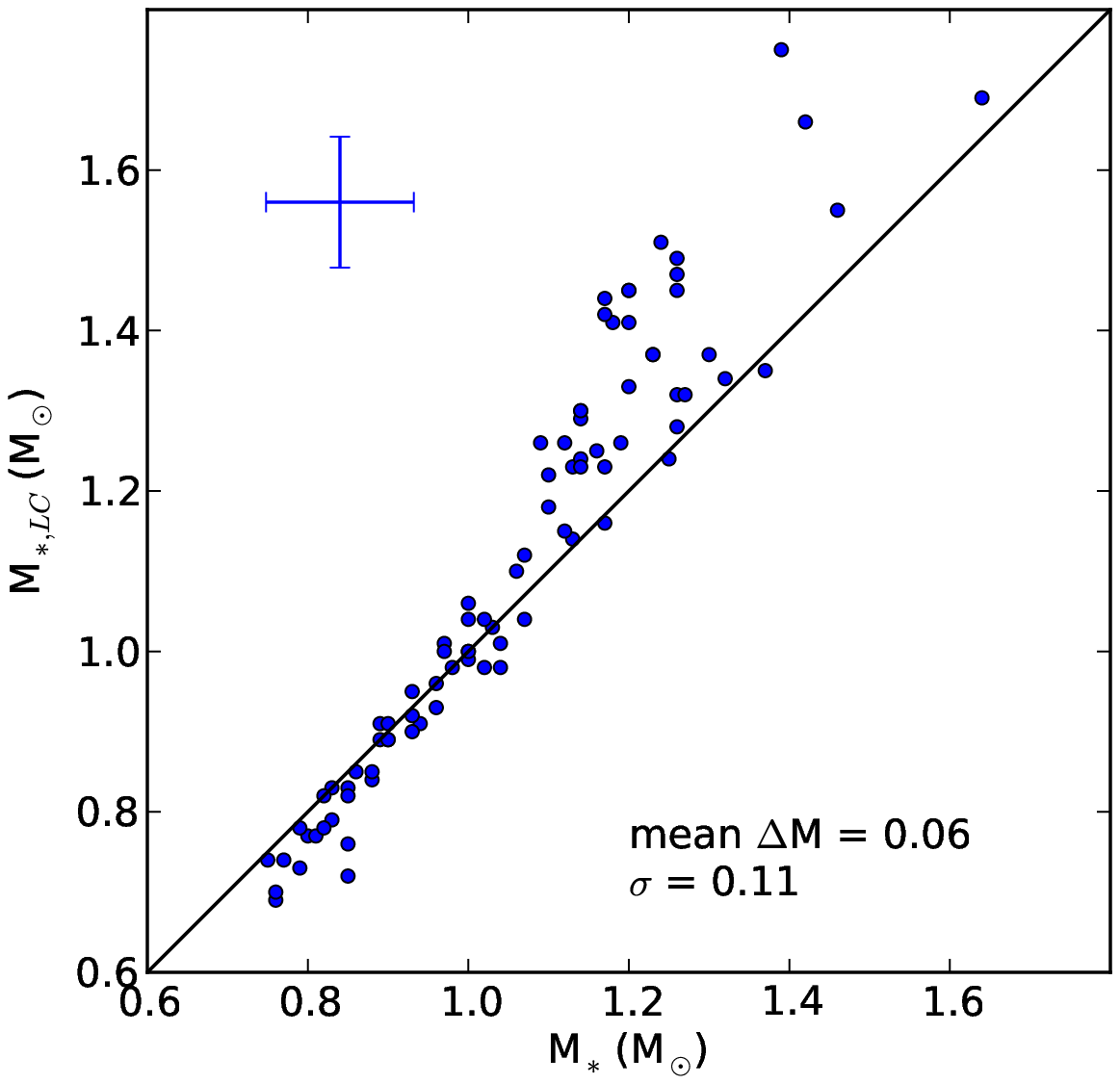}
\includegraphics[width=5.8cm]{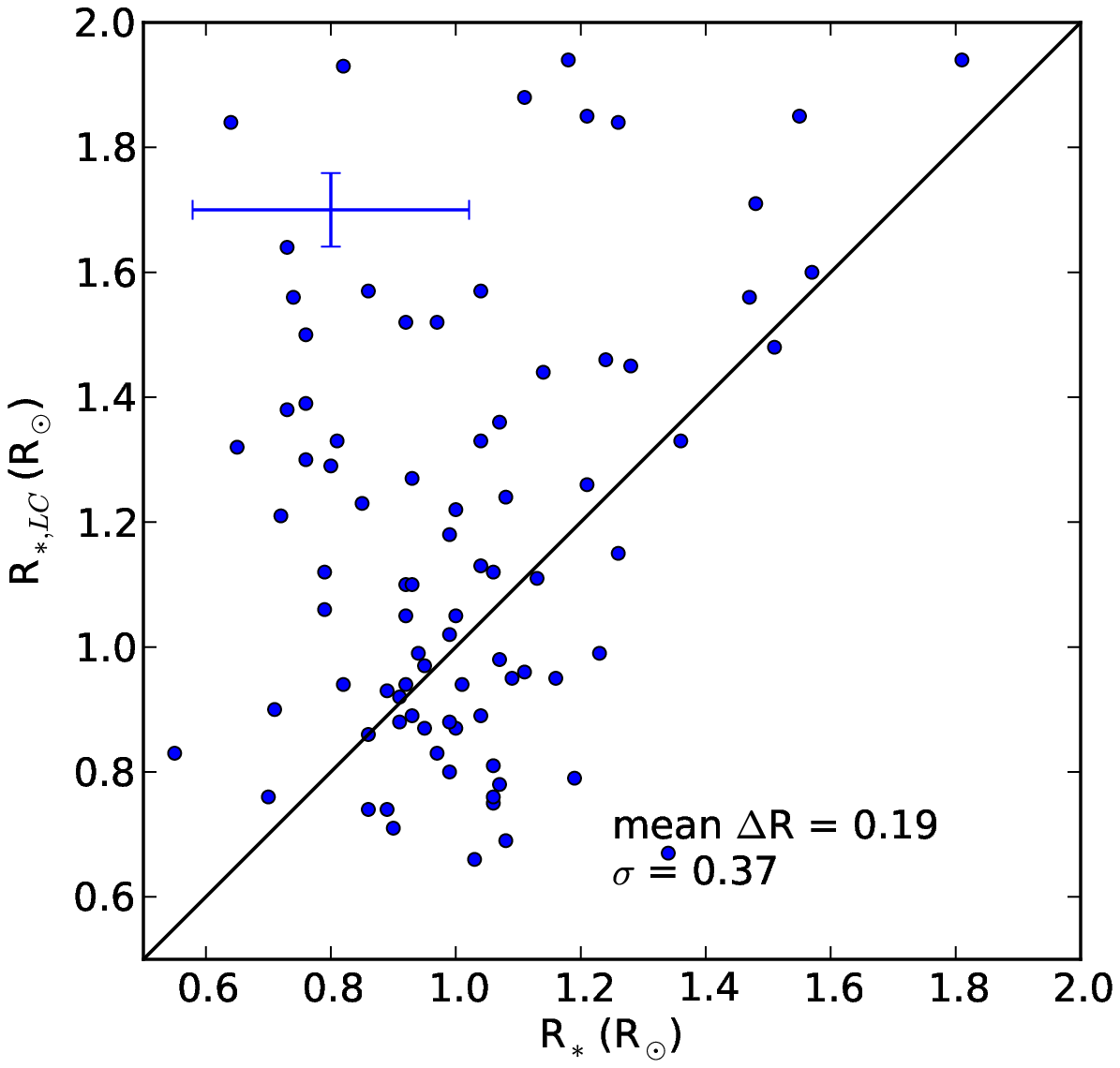}
\includegraphics[width=5.8cm]{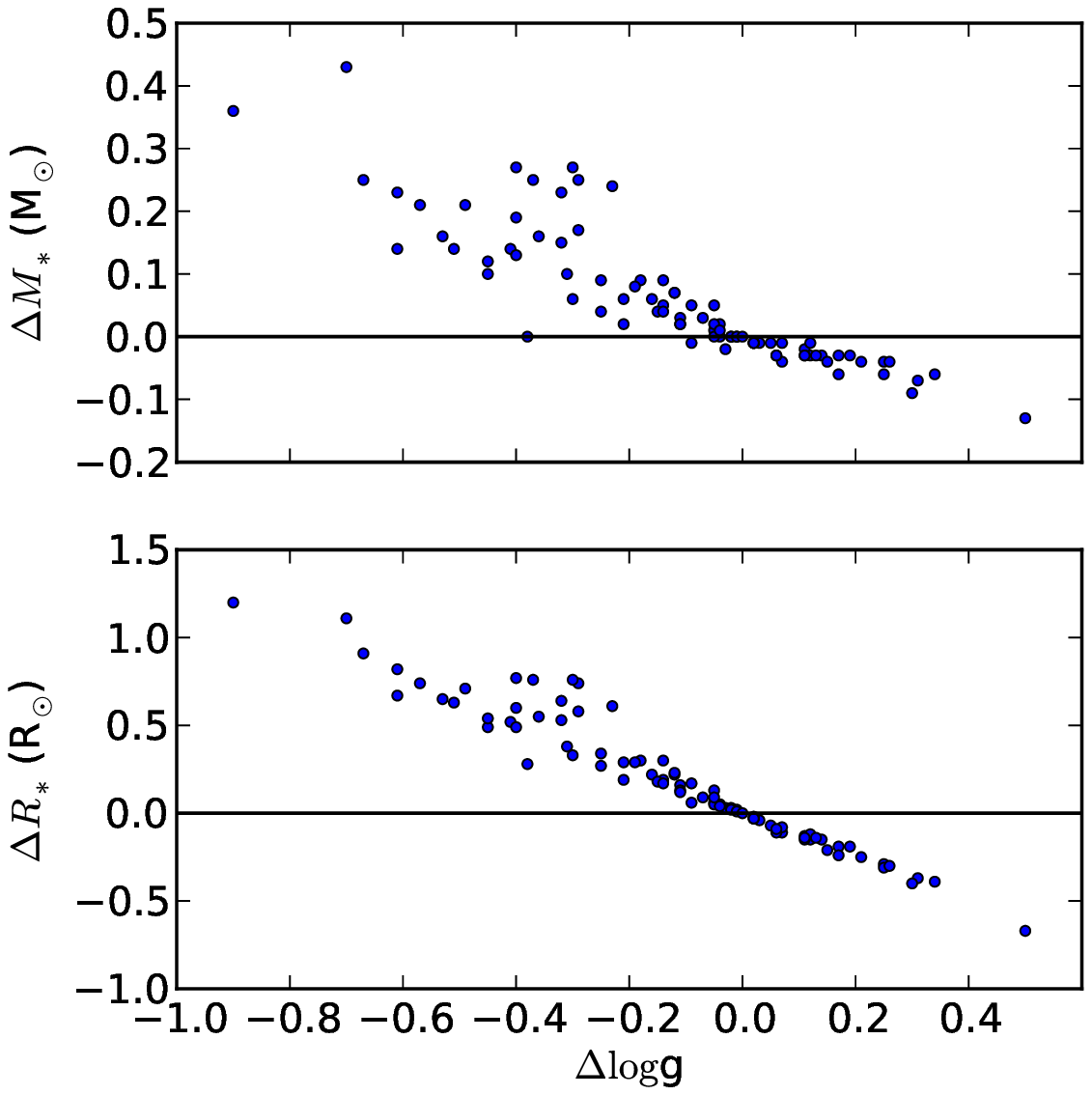}
\caption{Comparisons of the masses and radii, obtained with the spectroscopic and photometric surface gravity. The right panel shows the differences as a function of the difference in surface gravity (all defined as `photometric - spectroscopic').}
\label{FigTor}
\end{center}
\end{figure*}

With the new photometric surface gravity, we also recalculated the mass and radius of every star, using the calibrations from \citet{Tor10}. Results are listed in Table \ref{TabPar}. Figure \ref{FigTor} shows the comparisons between these values. The masses compare well with a mean difference of $0.06$\,M$_{\odot}$. The greatest differences are found for higher mass stars. The radii, on the other hand, do not compare so well. In the righthand panel of Figure \ref{FigTor}, we plot the differences in masses and radii with respect to the surface gravity difference (all defined as `photometric - spectroscopic $\log$g'). Clear linear trends are visible.

For the radii, the effect of using different surface gravities is greatest with absolute differences up to $1.0$\,R$_{\odot}$. These large discrepancies in stellar radii can lead to large discrepancies in planetary radii (see Section \ref{Disc}). Since the photometric surface gravity is generally more precise than the spectroscopic one, the resulting stellar masses and radii will also be more precise.

\section{Comparison with the literature}\label{Lit}

\begin{figure*}[t!]
\begin{center}
\includegraphics[width=5.8cm]{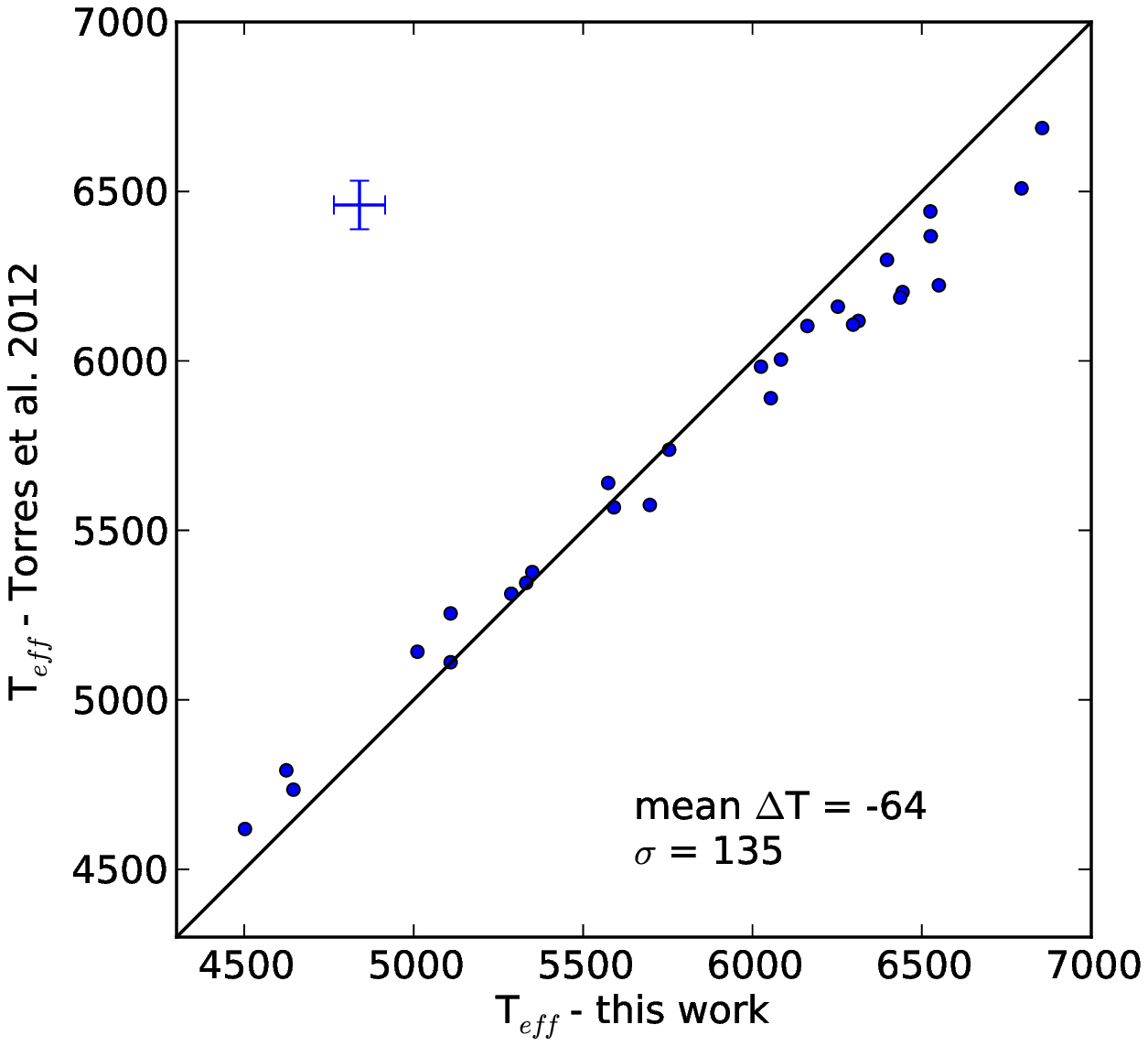}
\includegraphics[width=5.8cm]{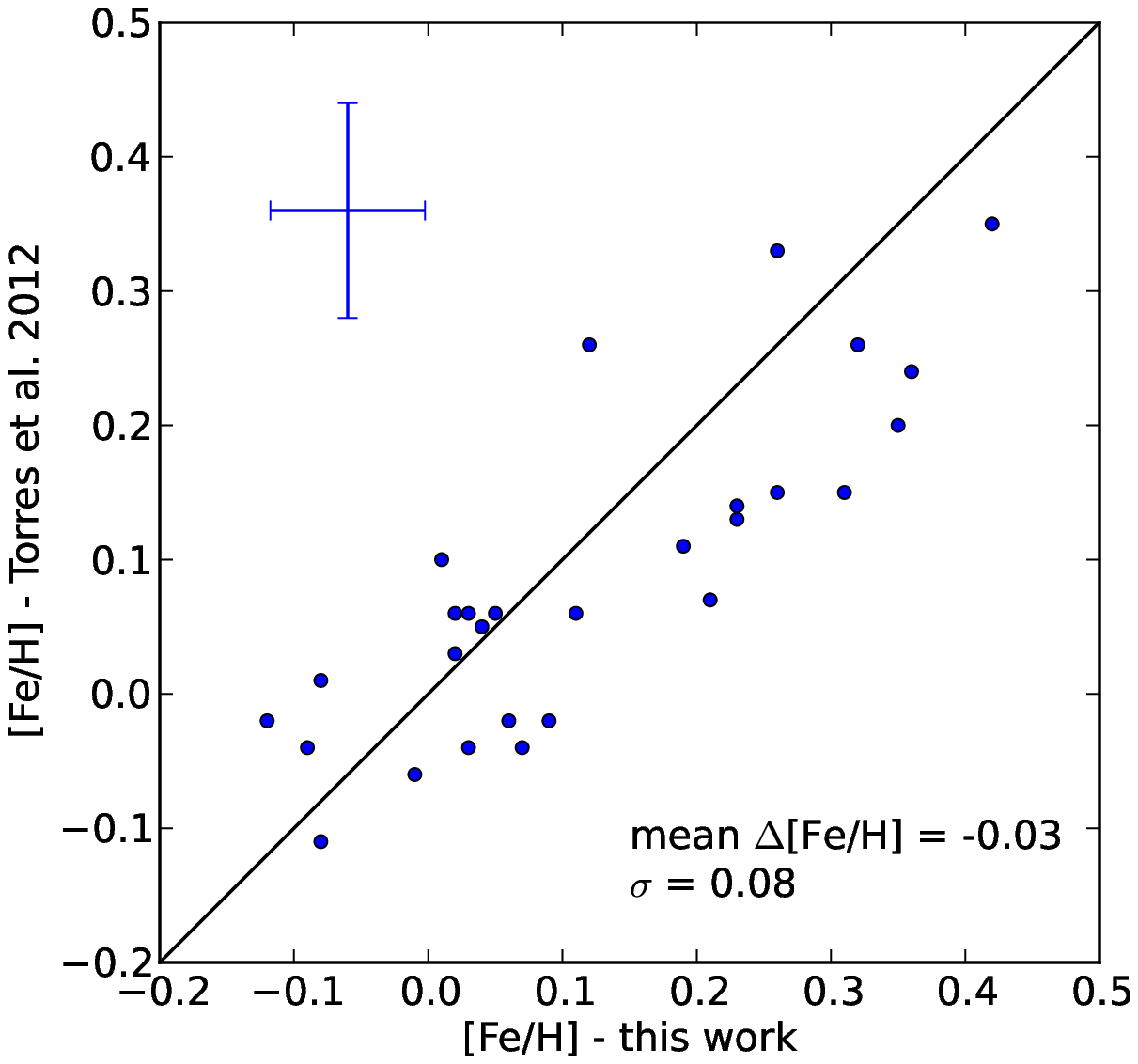}
\includegraphics[width=5.8cm]{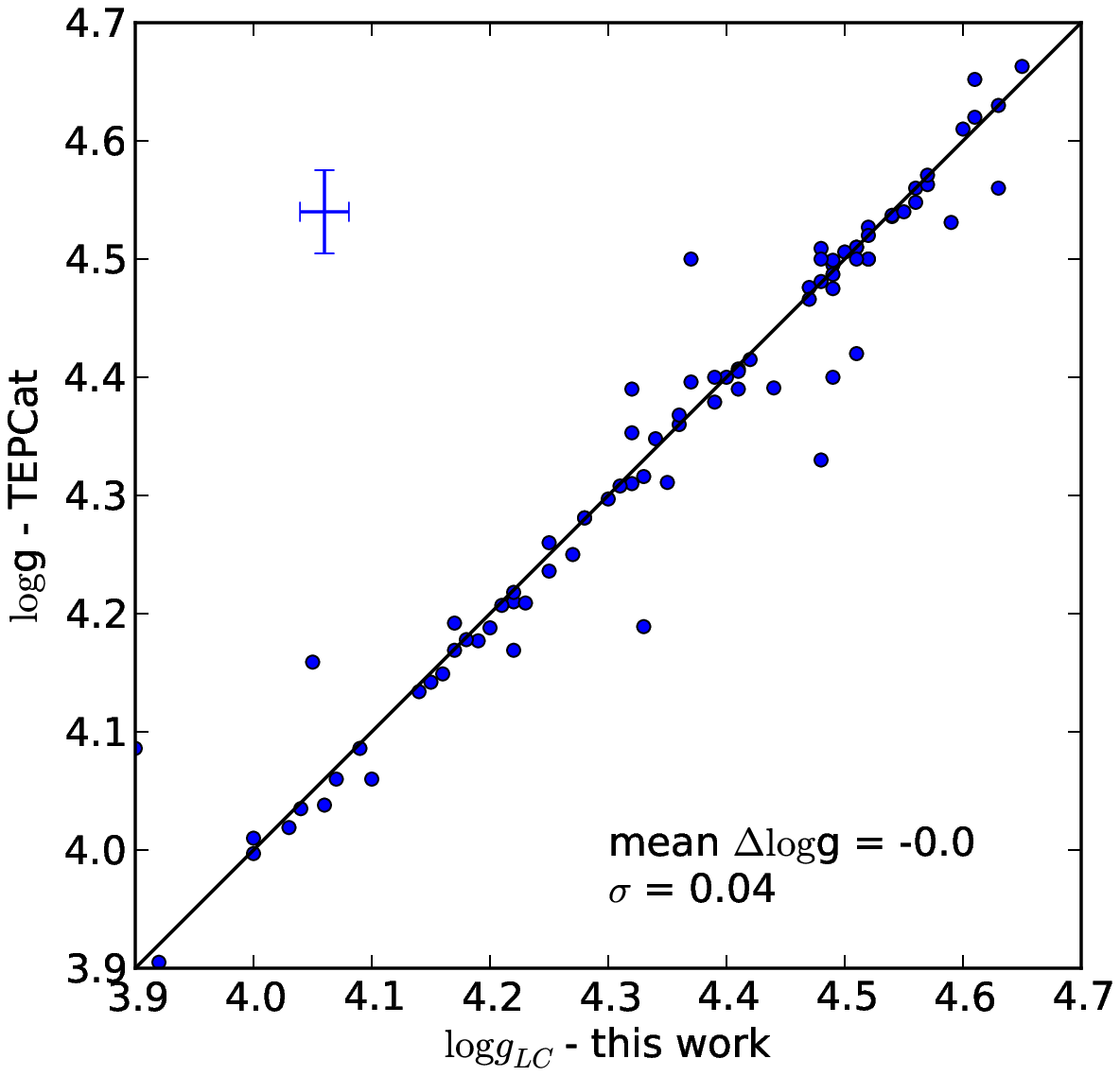}\\
\includegraphics[width=5.8cm]{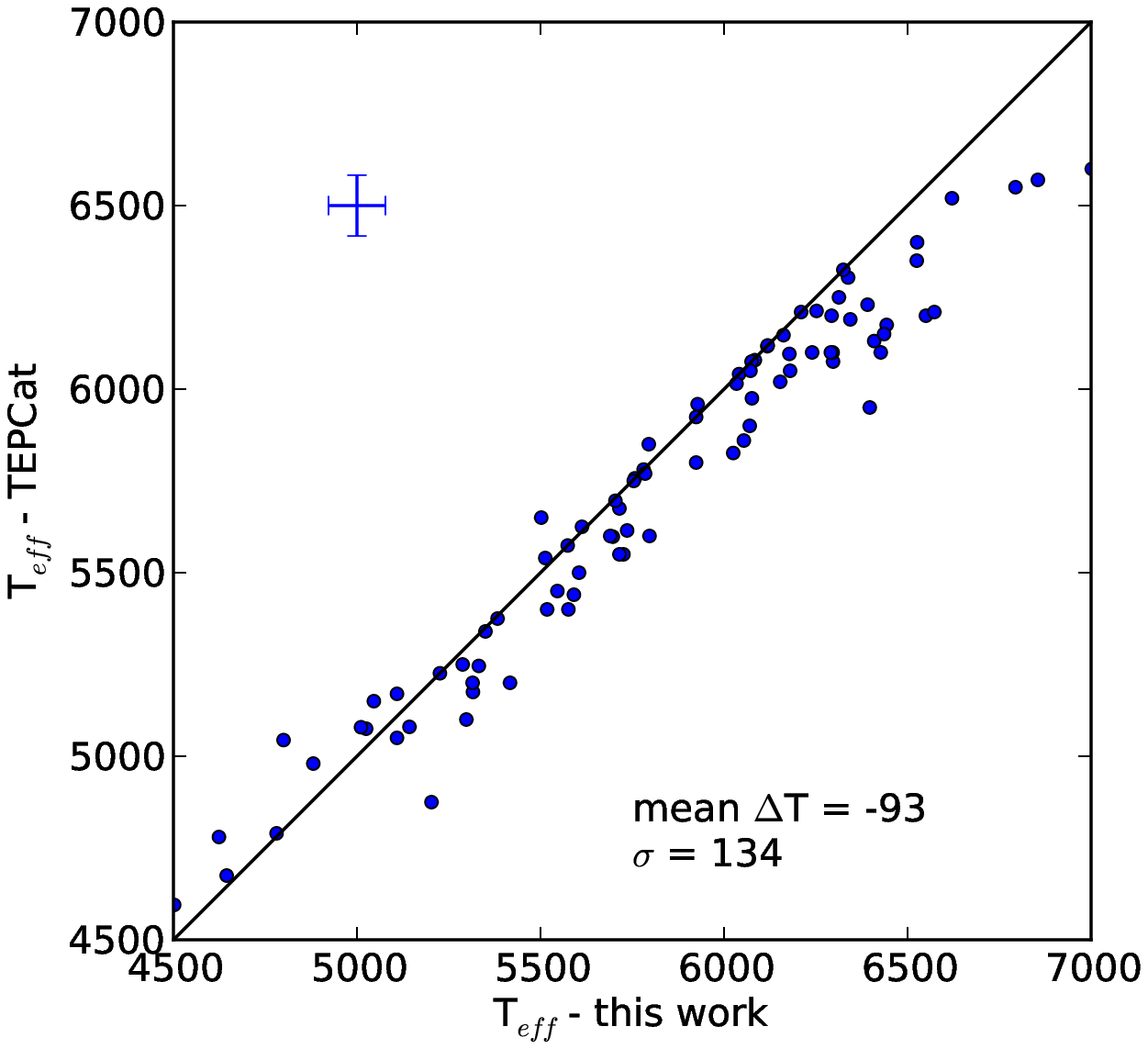}
\includegraphics[width=5.8cm]{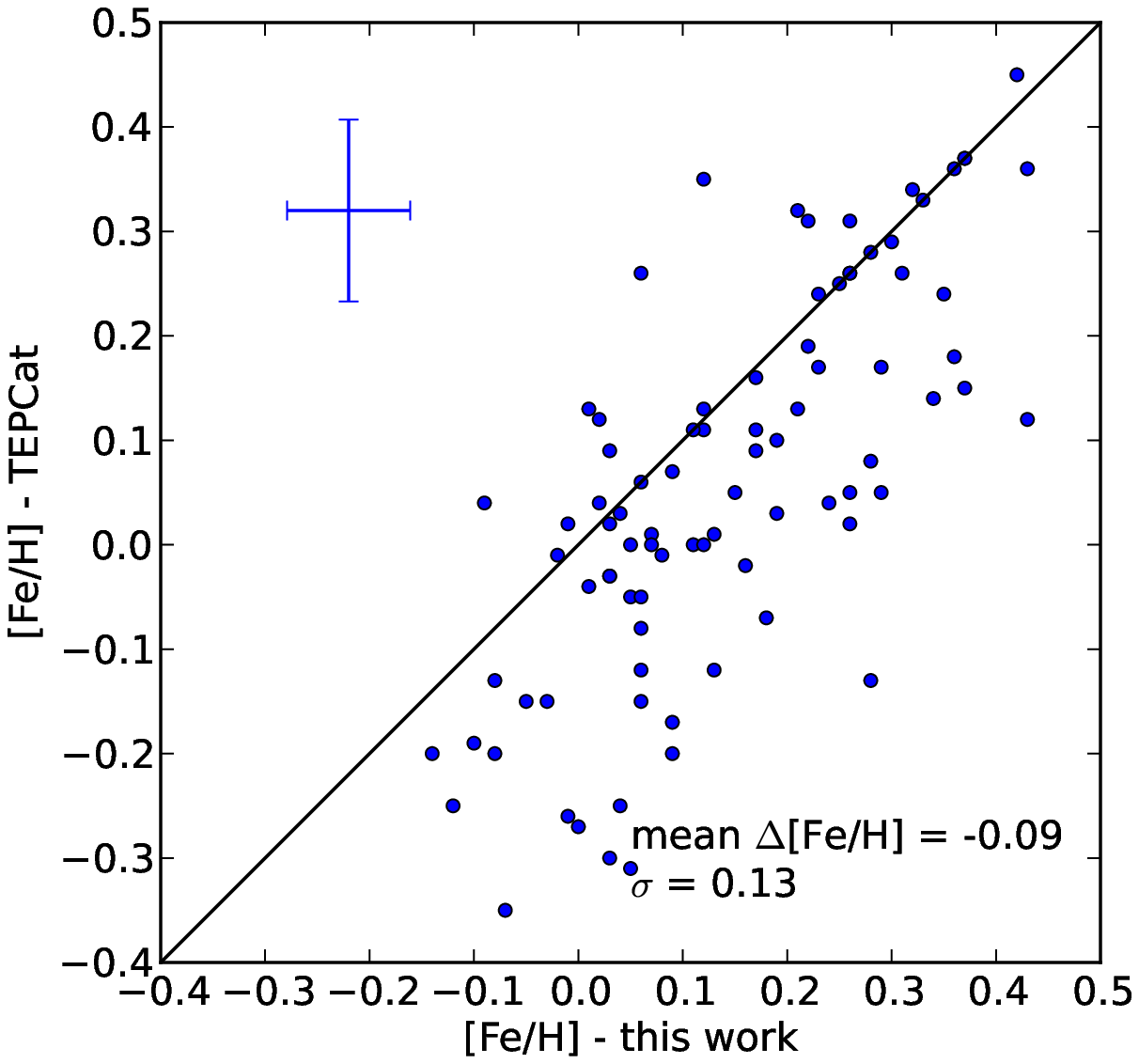}
\includegraphics[width=5.8cm]{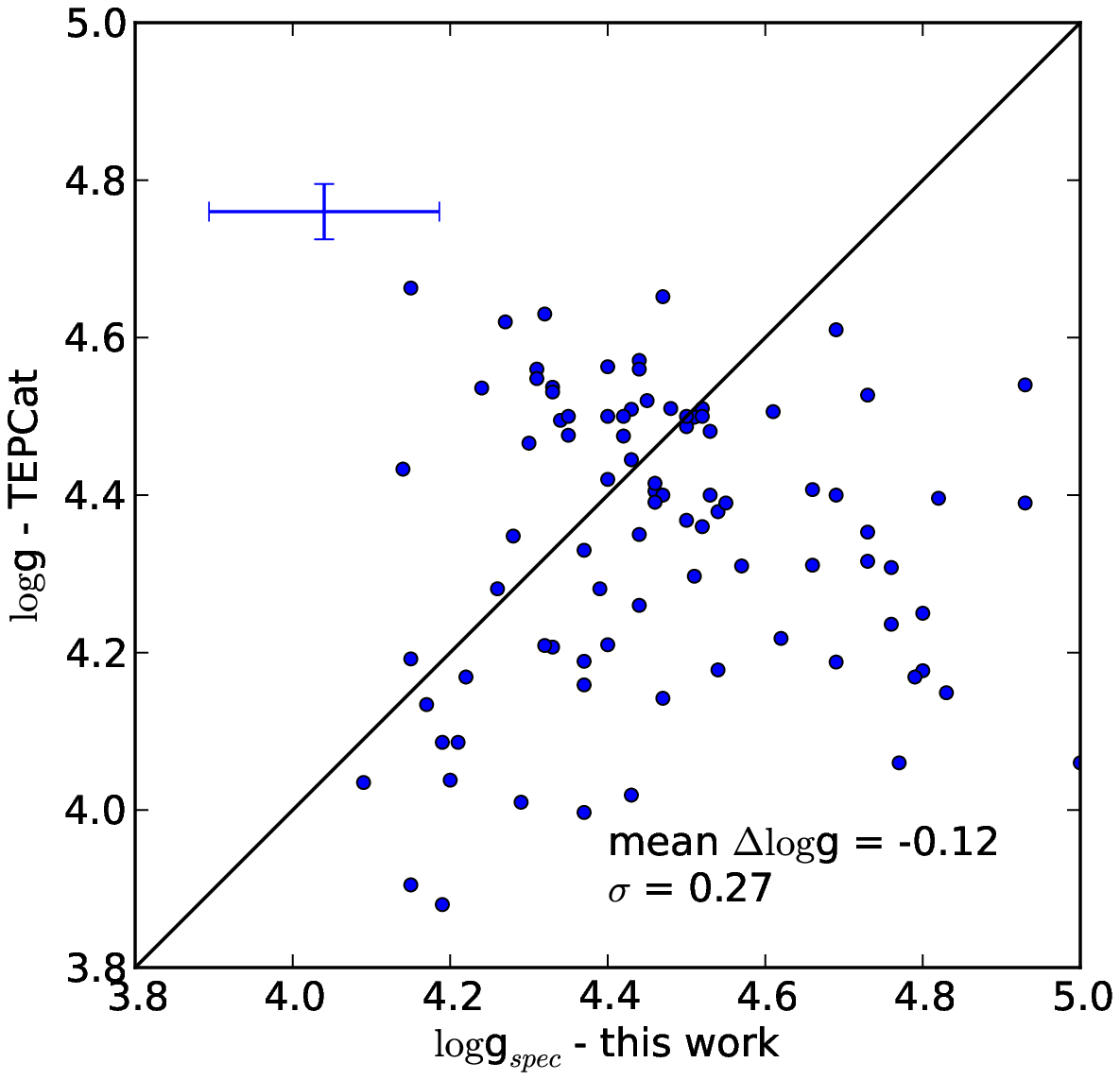}
\caption{Comparisons of the spectroscopic results in this work with the results of \citet{Tor12} (top left and midddle panel) and the results of the TEPCat catalog (bottom panels).}
\label{FigLit}
\end{center}
\end{figure*}

Recently, another homogeneous spectroscopic analysis has been done for transiting planet hosts by \citet{Tor12}. Their analysis of the temperature and metallicity is based primarily on the spectral classification technique, as described in \citet{Buch12}. They also use the spectroscopy made easy (SME) technique \citep{Val05} and MOOG. We have 28 stars in common with their sample. The comparisons are shown in the top panels of Figure \ref{FigLit}. Both the temperature and the metallicity compare well with a mean difference of $-64$\,K and $-0.03$\,dex, respectively. For the effective temperature, a slight deviation for higher temperatures can be seen. We do not compare with their surface gravities since they have taken them from external sources.

We also compared with all the values listed in the TEPCat catalog \citep{South11} where we have 88 stars in common. In the bottom panels we compare our spectroscopic results for the effective temperature, metallicity, and surface gravity with their results. The temperatures compare well, with a mean difference of $-93$\,K. The same slight deviation for higher temperatures can be seen. The metallicities show a mean difference of $-0.09$. There is also a wide spread present in this comparison. This shows again that a homogeneous analysis of stellar parameters is very important. As expected it can be seen that the spectroscopic surface gravities do not compare well. In the top righthand panel, we compare our light curve surface gravities with the values in the TEPCat catalog. It is immediately clear that these compare extremely well. On average, there is no difference between these surface gravities.

\section{Discussion}\label{Disc}

We found that stellar masses and radii are affected by using different surface gravities. Especially for stellar radii, the differences can go up to 1.0 R$_{\odot}$. Planetary radii are linearly affected by the stellar radius (the transit depth provides the radius ratio $R_p/R_{\ast}$). Caution should thus be placed when calculating planetary radii. 

\begin{figure}[t!]
\begin{center}
\includegraphics[width=8.cm]{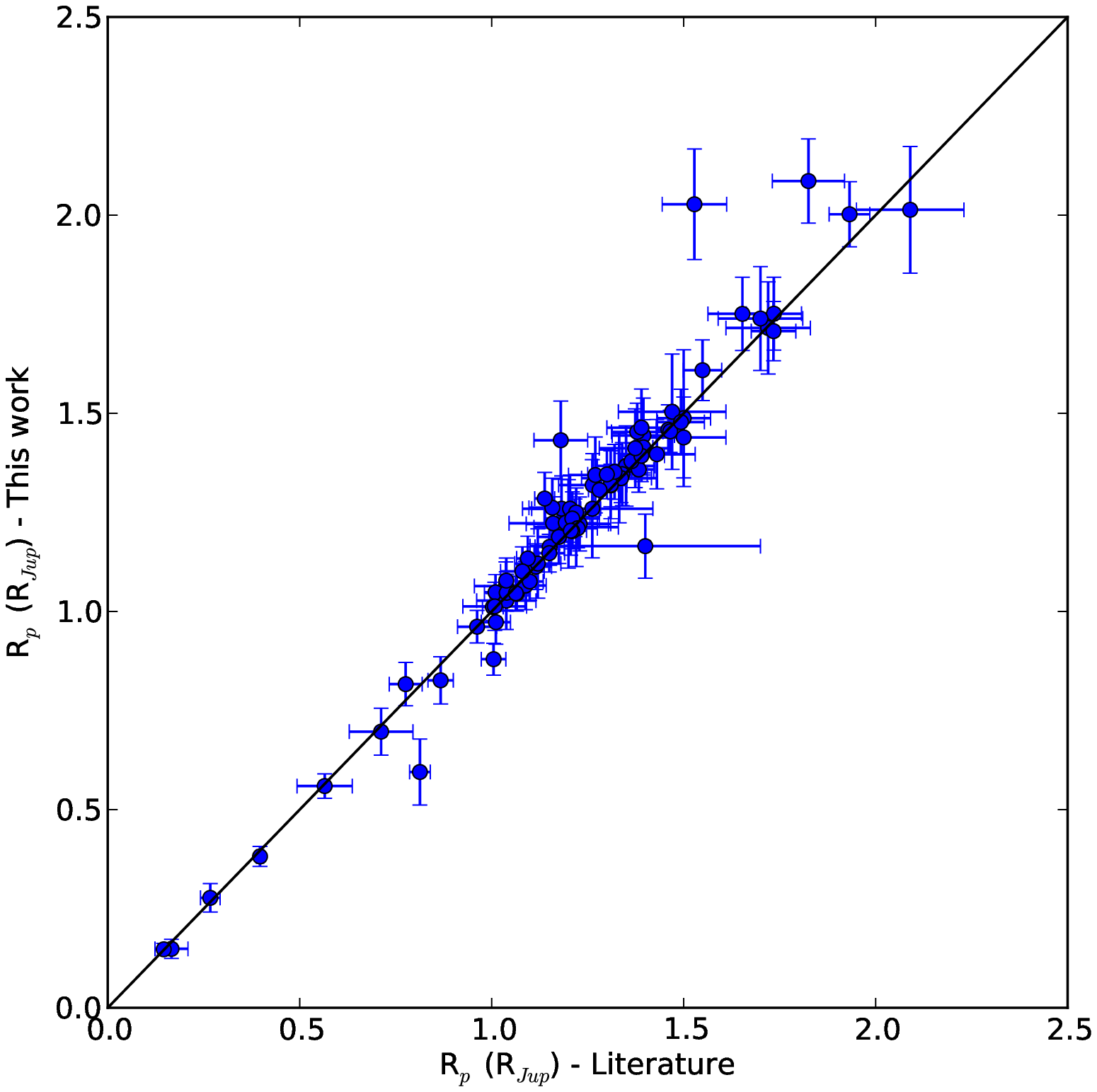}
\includegraphics[width=8.cm]{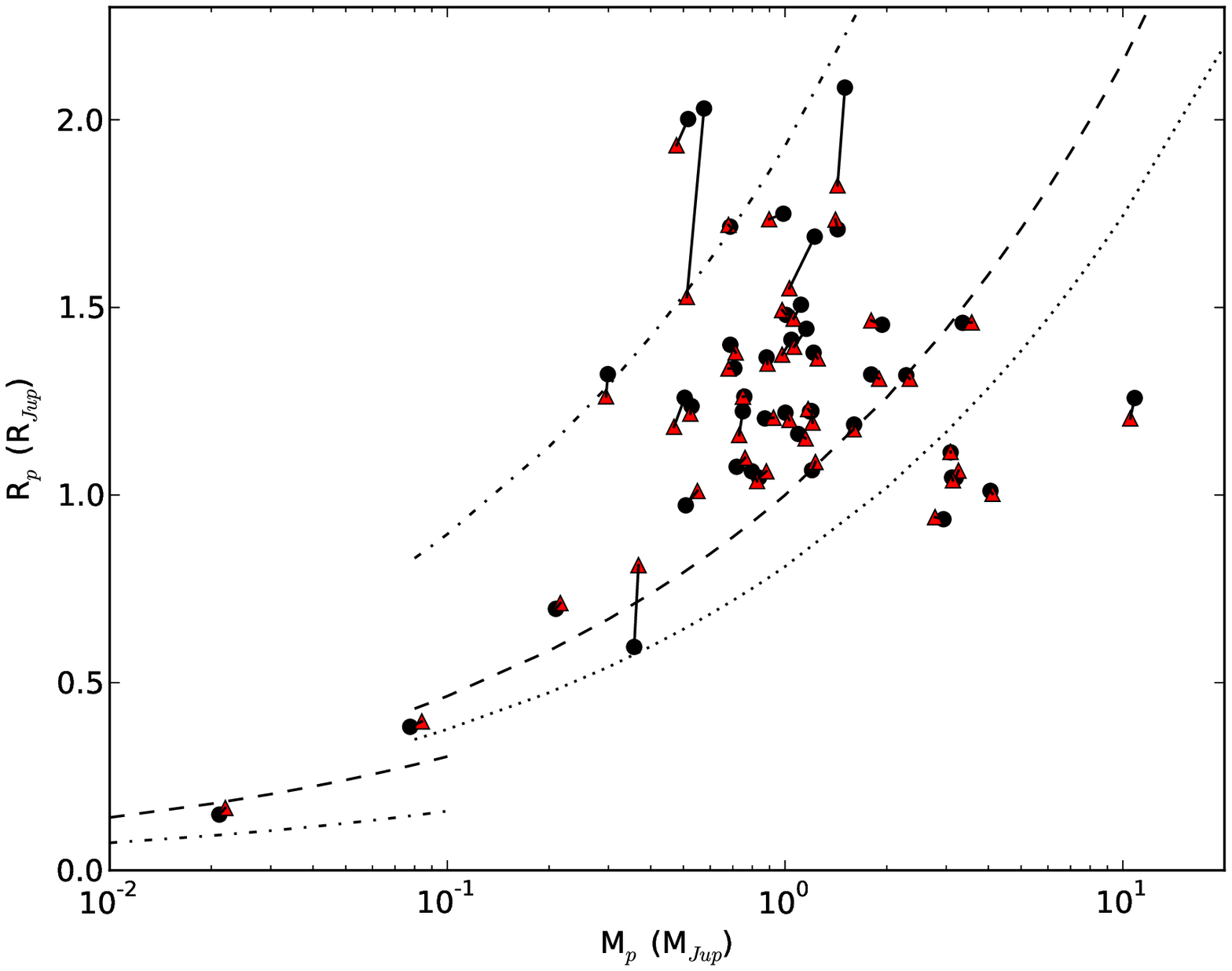}
\caption{Top panel: Comparison of the planetary radii, calculated with our photometric stellar radius, with literature values. Bottom panel: Planetary radii are plotted against their masses. Circles denote the values calculated with the stellar radius and mass from this work, using the photometric surface gravity, while the triangles are the values from the homogenous TEPCat Catalogue. Isodensity curves are overplotted for Saturn (dash-dotted), Jupiter (dashed), Neptune (dotted), Mars (dash-dotted), and Pluto (dashed).}
\label{FigPlanet}
\end{center}
\end{figure}

With our stellar radii, we recalculated all planetary radii for the planets from this sample. We used the radius ratios from the same works we used to get the stellar densities. The top panel of Figure \ref{FigPlanet} compares the new planetary radii calculated with our photometric stellar radius with the planetary radii from the literature works. Most planetary radii, especially the small ones, agree very well, within one sigma. Since most transit discovery papers calculate stellar radii based on a photometric surface gravity, this could be expected. However, there are still several planets where the difference in radius is more than two sigma (\object{CoRoT-1}, \object{HD 149026}, \object{WASP-11}, \object{WASP-12}, \object{WASP-13}, \object{WASP-32}, \object{WASP-50}, \object{WASP-8}). If one used stellar radii, which are calculated with spectroscopic surface gravities, the differences would be much greater. 

For the planet hosts that we have in common with the homogeneous part of the TEPCat Catalogue \citep{South10}, we also recalculated the planetary masses using our photometric stellar masses. In the bottom panel of Figure \ref{FigPlanet}, we plot the planetary radius versus the planetary mass. We use both our newly calculated values and the values from the TEPCat catalog. Since the stellar radius is more affected than the stellar mass by using a different surface gravity, the planetary radius is also more affected than the planetary mass. As already seen, most planetary values agree well, but for some planets, the radii differ a lot. This can influence theoretical composition models for these extrasolar planets. Overplotted in Figure \ref{FigPlanet} are isodensity curves for some planets from the solar system. A large difference in stellar and thus planetary radius can lead to incorrectly classifying of a planet. Caution should thus be used on planetary radius determinations since precise stellar radius determinations are very dependent on a precise determination of the atmospheric stellar parameters. In a forthcoming work, we will focus more on these planets for which we find very different parameters.

\section{Conclusions}\label{Con}

In this work, we spectroscopically derived stellar atmospheric parameters (effective temperature, surface gravity, metallicity and microturbulent velocity), stellar masses and radii, and chemical abundances for 90 transiting planet hosts, of which 28 were previously presented in works by members of our team. We used the ARES+MOOG method with carefully selected iron linelists. All values, calculated in this work, are added to the online SWEET-Cat catalog\footnote{\url{https://www.astro.up.pt/resources/sweet-cat/}} \citep{San13}.

We can summarize the results as follows.

\begin{itemize}
\item Temperatures and metallicities in general compare well with different literature sources.

\item Spectroscopically derived surface gravities are very poorly constrained. They were independently derived from the photometric light curve, using the spectroscopic temperatures and metallicities and stellar densities from the discovery papers. These new photometric surface gravities are much more precise and match, in general, the literature data very well. 

\item The chemical abundances were derived again using the photometric surface gravity. The abundances of the atoms are not affected by using different surface gravities. Abundances of ions, however, are slightly affected, as predicted by \citet{Gray92}.

\item Stellar masses and radii were derived through calibration formulae based on the effective temperature, metallicity, and surface gravity. The different values of the surface gravity do not have any strong effect on the mass determination with only a mean difference of 0.06M$_{\odot}$, but it does on the radius determination where the comparison shows a large spread. Using the more precise photometric surface gravity also results in more precise stellar mass and radius determinations. 

\item Planetary radii and masses were recalculated using the photometric stellar radii and masses. Most values agree within error bars, but eight planets show differences of more than two sigma.

%\item Using stellar radii, based on spectroscopic surface gravities, will provide less precise and less accurate planetary radii.
\end{itemize}

\begin{acknowledgements}
      This work made use of the ESO archive and the Simbad Database. This work was supported by the European Research Council/European Community under the FP7 through Starting Grant agreement number 239953. N.C.S. acknowledges the support of the Fundação para a Ciência e a Tecnologia (FCT) in the form of grant reference PTDC/CTE-AST/098528/2008. V.Zh.A., S.G.S. and E.D.M are supported by grants SFRH/BPD/70574/2010, SFRH/BPD/47611/2008 and SFRH/BPD/76606/2011, respectively , also from FCT. GI acknowledges financial support from the Spanish Ministry project MICINN AYA2011-29060.

\end{acknowledgements}

\bibliographystyle{aa} 
\bibliography{/home/annelies/My_Articles/References.bib}

\end{document}